\documentclass[fleqn,usenatbib]{mnras}
\usepackage{mathptmx}
\usepackage[T1]{fontenc}

\DeclareRobustCommand{\VAN}[3]{#2}
\let\VANthebibliography\thebibliography
\def\thebibliography{\DeclareRobustCommand{\VAN}[3]{##3}\VANthebibliography}

\usepackage{xspace}
\usepackage[colorinlistoftodos]{todonotes}

\usepackage{graphicx}	
\usepackage{amsmath}	
\usepackage{amssymb}	
\usepackage{siunitx}

\title[Vertical structure of NGC~3501]{The vertical structure of the spiral galaxy NGC 3501: first stages of the formation of a thin metal-rich disc}

\author[N. Sattler et al.]{Natascha Sattler$^{1}$\thanks{E-mail: sattler@mpia.de}, Francesca Pinna$^{1}$, Nadine Neumayer$^{1}$, Jesus Falc\'on-Barroso$^{2,3}$, Marie Martig$^{4}$, \newauthor Dimitri A. Gadotti$^{5,6}$, Glenn van de Ven$^{7}$, Ivan Minchev$^{8}$
\\
$^{1}$Max-Planck-Institut für Astronomie, Königstuhl 17, D-69117 Heidelberg, Germany\\
$^{2}$Instituto de Astrof{\'i}sica de Canarias, Calle Via L{\'a}ctea s/n, 38200 La Laguna, Tenerife, Spain\\
$^{3}$Depto. Astrof{\'i}sica, Universidad de La Laguna, Calle Astrof{\'i}sico Francisco S{\'a}nchez s/n, 38206 La Laguna, Tenerife, Spain\\
$^{4}$Astrophysics Research Institute, Liverpool John Moores University, 146 Brownlow Hill, Liverpool L3 5RF, UK\\
$^{5}$European Southern Observatory, Karl-Schwarzschild-Str. 2, D-85748 Garching, Germany\\
$^{6}$Centre for Extragalactic Astronomy, Department of Physics, Durham University, South Road, Durham DH1 3LE, UK\\
$^{7}$Department of Astrophysics, University of Vienna, Türkenschanzstraße 17, 1180 Vienna, Austria\\
$^{8}$Leibniz Institut fur Astrophysik Potsdam (AIP), An der Sterwarte 16, D-14482 Potsdam, Germany}

\date{Accepted XXX. Received YYY; in original form ZZZ}

\pubyear{2022}

\begin{document}
\label{firstpage}
\pagerange{\pageref{firstpage}--\pageref{lastpage}}
\maketitle


\begin{abstract}
We trace the evolution of the edge-on spiral galaxy NGC~3501, making use of its stellar populations extracted from deep integral-field spectroscopy MUSE observations. We present stellar kinematic and population maps, as well as the star formation history, of the south-western half of the galaxy.
The derived maps of the stellar line-of-sight velocity and velocity dispersion are quite regular, show disc-like rotation, and no other structural component of the galaxy.
However, maps of the stellar populations exhibit structures in the mass-weighted and light-weighted age, total metallicity and [Mg/Fe] abundance.
These maps indicate that NGC~3501 is a young galaxy, consisting mostly of stars with ages between 2 to 8 Gyr.
Also, they show a thicker more extended structure that is metal-poor and $\alpha$-rich, and another inner metal-rich and  $\alpha$-poor one with smaller radial extension.
While previous studies revealed that NGC~3501 shows only one morphological disc component in its vertical structure, we divided the galaxy into two regions: an inner metal-rich midplane and a metal-poor thicker envelope. 
Comparing the star formation history of the inner thinner metal-rich disc and the thicker metal-poor disc, 
we see that the metal-rich component evolved more steadily, while the metal-poor one experienced several bursts of star formation.
We propose this spiral galaxy is being observed in an early evolutionary phase, with a thicker disc already in place and an inner thin disc in an early formation stage. So we are probably witnessing the birth of a future massive thin disc, continuously growing embedded in a preexisting thicker disc.\\
\end{abstract}





\section{Introduction}

The vertical structure of external galaxies can be best studied when they are seen edge-on, where disc structures with different thicknesses can be identified. Thick and thin discs in external galaxies were already discovered and described many decades ago by \citet{burstein_1979}, and in the Milky Way by \cite{gilmore_1983}. Thick and thin discs have since then been identified in many more galaxies. For instance, \citet{yoachim_2006} analyzed the vertical structure of 34 edge-on galaxies and found that 32 of them host a thick disc.
In a further study of 141 galaxies by \citet{comeron_2018}, 17 galaxies show only one distinct disc component, while 124 galaxies host at least two large-scale disc components, like a thin and thick disc.

In external galaxies, thick and thin discs are usually defined geometrically, after fitting vertical surface-brightness profiles as done in \citet{comeron_2018}, or selecting stars far away from the mid-plane \citep[e.g.,][]{yoachim_2008a}. In this paper, we use this geometric definition of the thick disc: this is a component with a large scale-height, made of stars with large excursions above the midplane, and high velocity dispersion. Such thick discs are usually significantly fainter and less massive than thin discs \citep{yoachim_2006, comeron_2018}.

Thick and thin discs also generally differ in their stellar populations: thick discs typically consist of stars that are older, more metal-poor, and more enhanced in $\alpha$-elements than thin discs \citep[e.g.,][]{yoachim_2008b, comeron_2015, kasparova_2016, pinna_2019b, martig_2021, scott_2021}.
However, thick and thin discs do not necessarily have strong differences in age, as can be seen in some lenticular galaxies in galaxy clusters \citep{comeron_2016, pinna_2019a}. 
For the Milky Way, while in the solar neighborhood it is true that stars at higher distances from the midplane are older, more metal-poor and $\alpha$-rich \citep{gilmore_1985, ivezic_2008, schlesinger_2012, casagrande_2016}, this does not hold everywhere in the disc. The geometric thick disc does not have a strong radial metallicity gradient \citep{cheng_2012}, but has a clear gradient in [$\alpha$/Fe]: the outer thick disc mostly contains $\alpha$-poor stars \citep{hayden_2017, queiroz_2020, gaia-collaboration_2022}. A radial age gradient in the thick disc was also predicted by \cite{minchev_2015} and measured by \cite{martig_2016} with $\sim$~9~Gyr old stars in the inner part of the disc and $\sim$~5~Gyr old stars in the outer thick disc. 
Geometric thick discs can thus be complex components, reflecting the complexity and diversity in the star formation and assembly histories of their host galaxies.

There are two main pathways through which the stellar mass in a galaxy might be assembled: ''inside-out'' and ''outside-in'' formation \citep{sanchez-blazquez_2007, perez_2013}.
Galaxies with a mass larger than 10$^{10.5}$~M$_{\odot}$ \citep[e.g.,][]{perez_2013, pan_2015} grow from the ''inside-out'', meaning that the stellar mass is first assembled in the central midplane. 
This central star formation, as well as gas inflow to a central supermassive black hole, can then quench the star formation in the center by heating the cool gas from which new stars would be formed or by blowing it out \citep{schawinski_2014}.
The star formation is then shifted to the outer parts of the galaxy, where gas is accreted from the intergalactic medium or satellites.
This ''inside-out'' scenario would then result in a negative age gradient, where older stars are located in the central region and younger stars in the galaxy outskirts \citep{perez_2013}.
Low-mass galaxies in contrast are more likely to form from an ''outside-in'' process, resulting in a positive age gradient with younger stars in the center and older stars in the outer regions \citep{gallart_2008, zhang_2012}.

Thin discs, with an extended star formation and long chemical enrichment making them $\alpha$-poor and metal-rich, are thought to form mainly in-situ. Both gas-rich galaxy mergers and accretion of gas from the intergalactic medium might play a key role in providing large amounts of gas to fuel the extended and massive star formation in thin discs \citep{gallart_2019, martig_2021, conroy_2022}.

Thick discs themselves can also be explained by many different processes.
A thin disc can be dynamically heated by minor mergers \citep{quinn_1993, abadi_2003}, or the thick disc can be made up of stars accreted from satellite galaxies. 
Another possible mechanism to form a thicker disc is proposed by \citet{meng_2021}, where a rapid change in the orientation of the midplane leads to a thickening of the disc. 
Alternatively, thick discs could be born already thick, following violent disc instabilities \citep{bournaud_2009}, or because of one or more gas-rich mergers \citep{brook_2004}, where the turbulent gas carries lower angular momentum leading to the thicker disc shape. 

All these scenarios do not have to be mutually exclusive, and combinations of those were recently proposed \citep{pinna_2019a, pinna_2019b, martig_2021}. 
\cite{minchev_2015} also show that radial gradients in $\alpha$ abundance and age result from different formation channels for the inner and outer thick discs: the inner thick disc forms already thick at high redshift, while the outer thick disc progressively grows as younger populations are subjected to external perturbations. Those perturbations create some disc flaring and bring young, $\alpha$-poor stars into the outer regions of geometric thick discs.
Whether this picture proposed by \cite{minchev_2015} applies only to the Milky Way or to many other galaxies is still unclear: our general understanding is partly limited by the small amount of spectroscopic data available for external galaxies.


But obtaining these properties with spectroscopic observations of thick discs is challenging, because of the fast drop in surface brightness when moving away from the midplane. A few studies have gathered long-slit observations \citep{yoachim_2008a, yoachim_2008b, du_2017, katkov_2019, kasparova_2020} to measure the kinematic properties, as well as the stellar populations for thin and thick discs of several edge-on galaxies.

Nevertheless, using integral-field spectroscopy to continuously map stellar kinematics and populations is significantly more efficient.
The first dedicated study of an edge-on galaxy using integral-field units was done by \citet{comeron_2015} on ESO 533-4 with VIMOS, but with a relatively poor spatial resolution.
Then the first attempt using MUSE (Multi-Unit Spectroscopic Explorer) data was performed by \citet{comeron_2016}, where the kinematics and stellar populations of a high-mass, edge-on S0 galaxy were studied.
So far, there are very few studies in the literature using integral field spectroscopy and most of them focus on bright S0 galaxies \citep{comeron_2016, guerou_2016, pinna_2019a, pinna_2019b}.
Due to low surface brightness, strong emission lines and dust, similar studies in late-type galaxies are significantly more challenging. 
Nevertheless \citet{comeron_2019} obtained spatial resolved maps of the stellar kinematics of 8 late-type galaxies, 
whereas \citet{martig_2021} and \citet{scott_2021} studied the spiral galaxies NGC~5746 and UGC 10738.\\

This paper extends this work by studying the stellar kinematics and populations, as well as the formation and evolution of the edge-on spiral galaxy NGC~3501. 
This galaxy was one out of the 17 galaxies from \citet{comeron_2018} whose vertical surface brightness profiles were fitted and showed only one disc component.
This paper is structured as follows: 
First, we will give a brief overview of the target galaxy in Section~\ref{sec:gal}, followed by a description of the observations and data reduction in Section~\ref{sec:obs} and the different analysis methods in Section~\ref{sec:ana}. 
Then we present the results in Section~\ref{sec:res}, discuss them in Section~\ref{sec:dis} and finally give our conclusion in Section~\ref{sec:con}.


\section{NGC 3501}
\label{sec:gal}

\begin{table}
	\centering
	\caption{Properties of NGC~3501.}
	\begin{tabular}{p{1.35cm} p{1.62cm} p{1.38cm} p{2.55cm}}
		\hline
		\hline
		\\
		\hfill{}Property & Value & Comments & Reference\\
		\hline
		\\
		\hfill{}Type & Scd & & \citet{ann_2015}\\
		\hfill{}RA & 11$^{h}$02$^{m}$47.307$^{s}$ & [J2000.0] & \citet{skrutskie_2006}\\
		\hfill{}Dec & +17$^{\circ}$59'22.31'' & [J2000.0] & \citet{skrutskie_2006}\\
    	\hfill{}PA & 28 & [$^{\circ}$] & \citet{comeron_2018}\\
    	\hfill{}Distance & 23.55 & [Mpc] & \citet{tully_2016}\\
		\hfill{}Redshift & 0.00377 & & \citet{haynes_2018}\\
		\hfill{}Major-axis & 233.4 & D$_{25}$ [arcsec] & \citet{de-vaucouleurs_1991}\\
    	\hfill{}Minor-axis & 30.81 & d$_{25}$ [arcsec] & \citet{de-vaucouleurs_1991}\\
            \hfill{}Thick disc scale height & 3.3 & [arcsec] & \citet{comeron_2018}\\
            \hfill{}Thin disc scale height & 1.0 & [arcsec] & \citet{comeron_2018}\\
    	\hfill{}Stellar~mass & 1.5 & [$\times$~10$^{10}$~M$_{\odot}$] & \citet{sheth_2010}\\
    	\hfill{}Dust mass & 1.1 & [$\times$~10$^{7}$~M$_{\odot}$] & \citet{nersesian_2019}\\
    	\hfill{}$v_c$ & 136 & [km s$^{-1}$] & \citet{sheth_2010}\\
		\hline
		\hline
		\\
	\end{tabular}
	\label{tab:prop}
\end{table}

\begin{figure}
    \includegraphics[width=1.\columnwidth ]{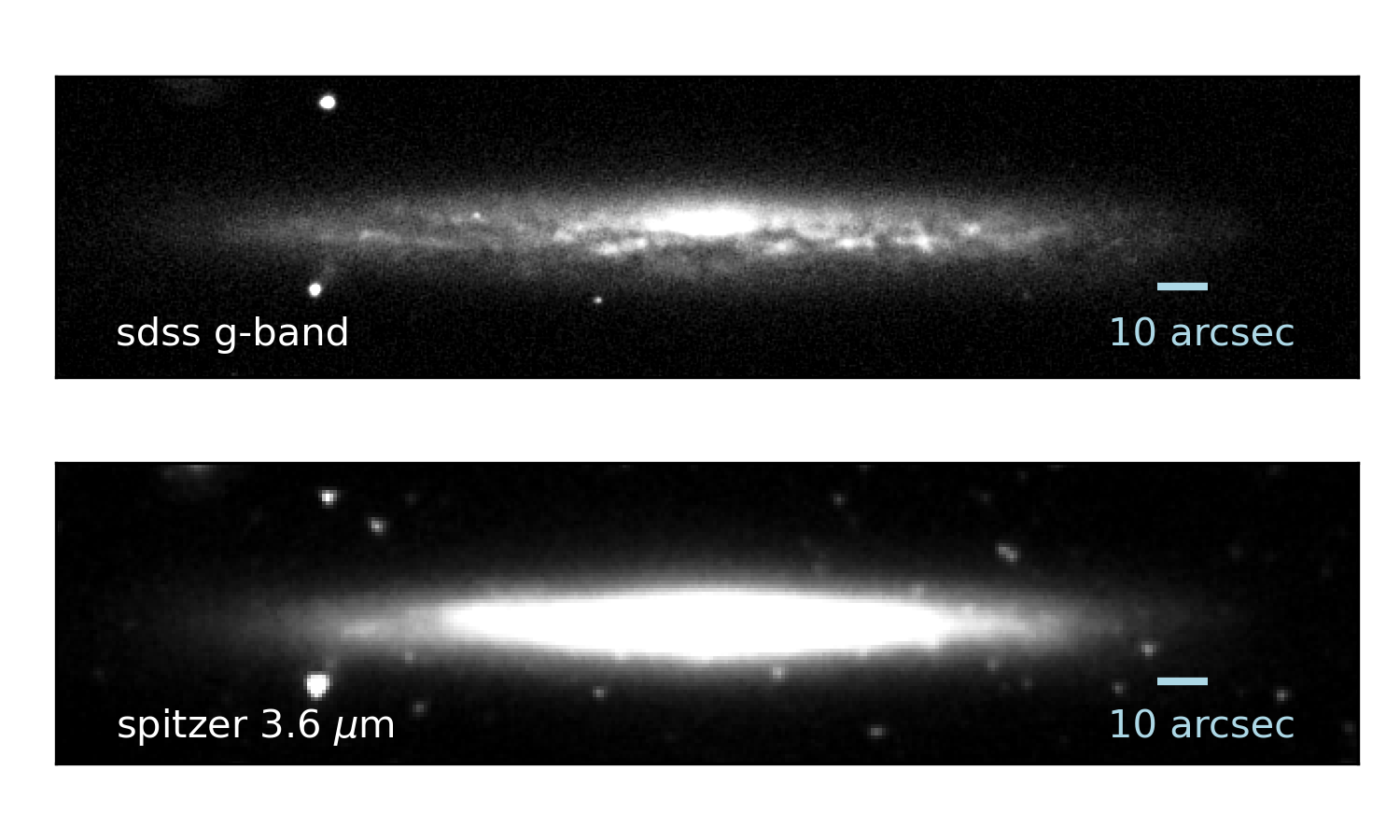} 
	\caption[]{SDSS g-band image \citep{ahumada_2020} of NGC~3501 at the top and S$^4$G $3.6 \mu$m image \citep{sheth_2010, munoz_2013, querejeta_2015} of NGC~3501 at the bottom.}
	\label{fig:sdss_s4g}
\end{figure}

NGC~3501 is an edge-on Scd galaxy with a boxy bulge \citep{luetticke_2000}. Some properties of NGC~3501 are listed in Table~\ref{tab:prop}, while Figure~\ref{fig:sdss_s4g} shows an SDSS g-band \citep{ahumada_2020} and an S$^4$G $3.6 \mu$m image \citep{sheth_2010, munoz_2013, querejeta_2015}.
This spiral galaxy is with 1.5~$\times$~10$^{10}$~M$_{\odot}$ slightly less massive than the Milky Way (5.43~$\times$~10$^{10}$~M$_{\odot}$, \citealt{mc-millan_2016}). It has a star formation rate of 0.52~M$_{\odot}$/yr \citep{nersesian_2019}, which is slightly below the Milky Way star formation rate of 0.68 to 1.45~M$_{\odot}$/yr \citep{robitaille_2010}.

Using 3.6~$\mu$m data from the S$^4$G survey \citep{sheth_2010}, \citet{comeron_2018} found that the vertical structure of the disc can be fitted by two components, a thin and a thick disc. However, the thick disc component always dominates for NGC~3501, so they proposed a single disc structure for this galaxy.
This single thicker disc of NGC~3501 has a fitted scale height of 3.3~arcsec (0.37~kpc), while the fitted scale height of a thin disc for NGC~3501 is 1.0~arcsec (0.11~kpc) \citep{comeron_2018}.
Therefore, the thicker disc of NGC~3501 is comparable in size to the Milky Ways thin disc, which has a scale height of 0.3~kpc \citep{juric_2008}. And it is significantly thinner than the Milky Ways thick disc, which has a scale height of 0.9~kpc \citep{juric_2008}.

\citet{marquez_2002} determined rotation curves and metallicity gradients using long-slit spectra in the region of H$\alpha$ emission. The rotation curve of NGC~3501 showed a maximum rotation velocity of around 100~km~s$^{-1}$ within 100~arcsec of the center with large errorbars of around 50~km~s$^{-1}$ in the outer regions.

\citet{marquez_2002} also found a negative radial metallicity gradient, as probed by [NII]/H$\alpha$ in HII regions. They also classified NGC~3501 as a mildly interacting spiral that shows non-disruptive interaction with satellites or companions.

\citet{marino_2010} imaged three nearby galaxy groups (LGG~93,
LGG~127 and LGG~225) in the far and near ultraviolet with
the Galaxy Evolution Explorer. NGC~3501 and its companion
NGC~3507 are members of LGG~225. They found for 9 spiral galaxies in LGG~225 luminosity-weighted ages from a few to around
7~Gyr. They also found that LGG~225 is in a pre-virial collapse phase and still undergoing dynamical relaxation. This means that galaxies in LGG~255 are in a more active evolution phase than the other two groups. Moreover, the two galaxies NGC~3501 and NGC~3507 are about 12.6~arcmin (86~kpc) apart, show no signs of interaction in the optical \citep{knapen_2014} but show signatures of distortion in the UV morphology \citep{marino_2010}. 
Comparing their masses, NGC~3507 has a stellar mass of about $1.3~\times~10^{10}$~M$_{\odot}$  \citep{sheth_2010} which is similar to the total stellar mass of NGC~3501 being around $1.5~\times~10^{10}$~M$_{\odot}$ \citep{sheth_2010}.

Finally, \citet{martin-navarro_2012} studied the surface brightness profile of NGC~3501 and found a first break at a radius of $\sim$~67~arcsec (7.6~kpc) and a further truncation at $\sim$~100~arcsec (11.36~kpc) from the center. They propose that the inner break might be related to a threshold in the ongoing star formation and that the truncation is more likely a real drop in the stellar mass density of the disc associated with the maximum angular momentum of the stars.

With the use of integral-field spectroscopy and especially by mapping the stellar population and star formation history, we gain a more detailed view and can search for small differences in the disc structure.


\section{Observations and data reduction}
\label{sec:obs}

The data was obtained using MUSE \citep{bacon_2010} at the Unit Telescope~4 of the Very Large Telescope, located in the Atacama Desert of northern Chile.
MUSE is a panoramic integral-field spectrograph that has a field of view of 1~squared arcmin in Wide-Field Mode (WFM) and a spatial sampling of 0.2~squared arcsec.
It can observe objects in a wavelength range of 4650-9300~\si{\angstrom} and has a mean spectral resolution of $R=3000$.

NGC~3501 was observed in Period~98, as part of the ESO program 098.B-0662 (PI: Pinna, F.), from December~30, 2016 to February~20, 2017. We used two MUSE-WFM pointings, an inner one that covers the central region of the galaxy, and an outer one towards the outer south-western part of the galaxy. 
The inner pointing was observed in one observing block with a total exposure time of 2200~s, while the outer pointing was observed in 8 observing blocks with a total exposure time of 17920~s. This results in a total exposure time of 20120~s ($\sim$~5.6~h) for both pointings. 
The data was reduced as described in \citet{gadotti_2019}, using the MUSE pipeline (version~1.6) from \citet{weilbacher_2012}. The two single pointings were initially combined into a mosaic, but we found significant offsets right at the edges of the pointings in the stellar population parameters. These offsets could result from different sky subtraction in the two pointings, or perhaps from the particular way we build the mosaic. Investigating this issue is beyond the scope of the paper, as we found that these offsets were greatly minimized when working with the two cubes separately. The results presented here thus use measurements from the individual cubes. Since the bins of the outer pointing have a better spatial resolution than the bins of the inner pointing in the overlapping region, due to the higher signal-to-noise ratio (SNR) per pixel (because of the longer exposure time), we use the values from the outer pointing for the overlapping region.

\begin{figure}
    \includegraphics[width=\columnwidth ]{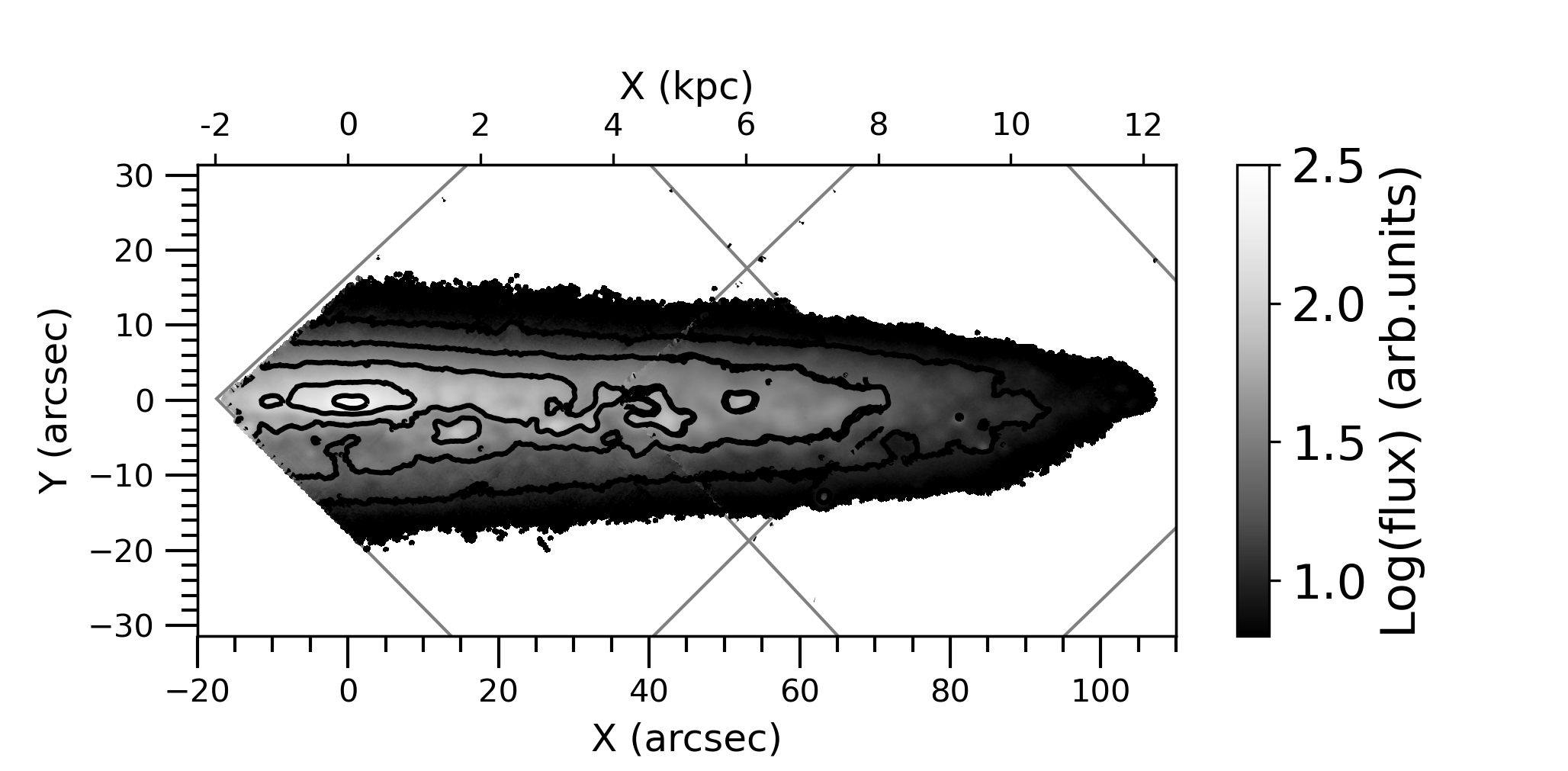} 
	\caption[]{Flux map of NGC3501 obtained with the two MUSE pointings. The approximate positions of the MUSE pointings are indicated by grey lines, while the isophotes are displayed in black}
	\label{fig:flux}
\end{figure}


\section{Analysis}
\label{sec:ana}

\subsection{Voronoi binning}
\label{subsec:binning}

The data cubes were binned, using the Voronoi binning method of \citet{cappellari_2003}. This method performs an adaptive spatial binning of integral-field spectroscopy data to reach a previously given minimum SNR per bin. It preserves the maximum possible spatial resolution of our data, for a given minimum SNR.

The resulting data of the inner pointing consists of 217 Voronoi bins with a target SNR~=~40 per bin, taking into account all spaxels. The outer pointing results in 130 Voronoi bins, also with a target SNR~=~40, but including only spaxels with a minimum SNR of 2. The different SNR thresholds were chosen such that the Voronoi bins of the two pointings extend to a similar flux level, and a continuous and clean transition at the edges of the two pointings is ensured.
In this step, the spectra of the bins were cut to the wavelength range of 4750 to 5500~\si{\angstrom}, so most regions in which the sky subtraction could have left residuals (at longer wavelengths) were avoided. Also, the age sensitivity of the used stellar population models (see Section~\ref{subsec:pop_models}) weakens towards longer wavelengths, which also include features that are not sensitive to $\alpha$-enhancement.
Finally, the data was cut to a specific spatial range to avoid residual spaxels appearing outside of the galaxy.
A flux map of NGC~3501 can be seen in Figure~\ref{fig:flux}.

Extending the wavelength range to around 7000~\si{\angstrom} and including the H$\alpha$ and [NII] emission lines is beyond the scope of this paper, and will be done in a future project.

\subsection{Stellar population models}
\label{subsec:pop_models}

In this paper we used MILES single-stellar population (SSP) models from \citet{vazdekis_2015}, which are based on BaSTI isochrones, to fit the spectra of the Voronoi bins.
All models have a spectral resolution of 2.51~\si{\angstrom} \citep{falcon-barroso_2011}, while covering the full wavelength range from 3540 to 7410~\si{\angstrom}. \citet{vazdekis_2015} also give safe ranges\footnote{http://research.iac.es/proyecto/miles/pages/ssp-models/safe-ranges.php} of SSP models for different spectral ranges and advise to use models with safe age and metallicity. Since we observed NGC~3501 in the optical and wanted to be in the safe range, we needed to exclude SSP models in metallicity and age. Therefore, we used models with 9 metallicity [M/H] values from $-1.26$ up to $+0.40$. Also, we used 28 age values going from 0.5 to 14~Gyr, evenly spaced every 0.5~Gyr, and two [$\alpha$/Fe] $\approx$ [Mg/Fe] values: 0.0~dex (solar abundance) and 0.4~dex (super-solar abundance). This results in a total amount of 504 models.

\subsection{Stellar kinematic fitting}
\label{subsec:kin_fit}

For the fitting of the stellar kinematics, we used the Python\footnote{www.python.org} version 7.4.2 of the full-spectrum penalized PiXel-Fitting (pPXF) method of \citet{cappellari_2004} and \citet{cappellari_2017}.
pPXF fits a combination of SSP models to the full stellar spectrum of the Voronoi bins, with a maximum penalized likelihood approach, to recover the line-of-sight velocity distribution (LOSVD). This is described as a Gauss-Hermite expansion and includes stellar mean velocity, velocity dispersion and higher moments. 
Since we observed that the galaxy has strong emission in H$\beta$ (4861.33~$\si{\angstrom}$), [OIII] (4958.92 \& 5006.84~$\si{\angstrom}$) and [NI] (5197.90 \& 5200.39~$\si{\angstrom}$), these spectral lines were masked with a masking window of 800~km~s$^{-1}$ and left out for the kinematic pPXF fitting.
We applied additive polynomials of 4$^{th}$ degree and used the first four moments ($V$, $\sigma$, $h_3$, $h_4$) of the LOSVD for the fitting, which were obtained as output. 
Also, we needed to discard 5 bins from the kinematic maps, because some spectra were too noisy or showed strong emission from foreground objects.

\begin{figure*}
	\begin{minipage}[l]{\columnwidth}
		\centering
		\includegraphics[width=\textwidth]{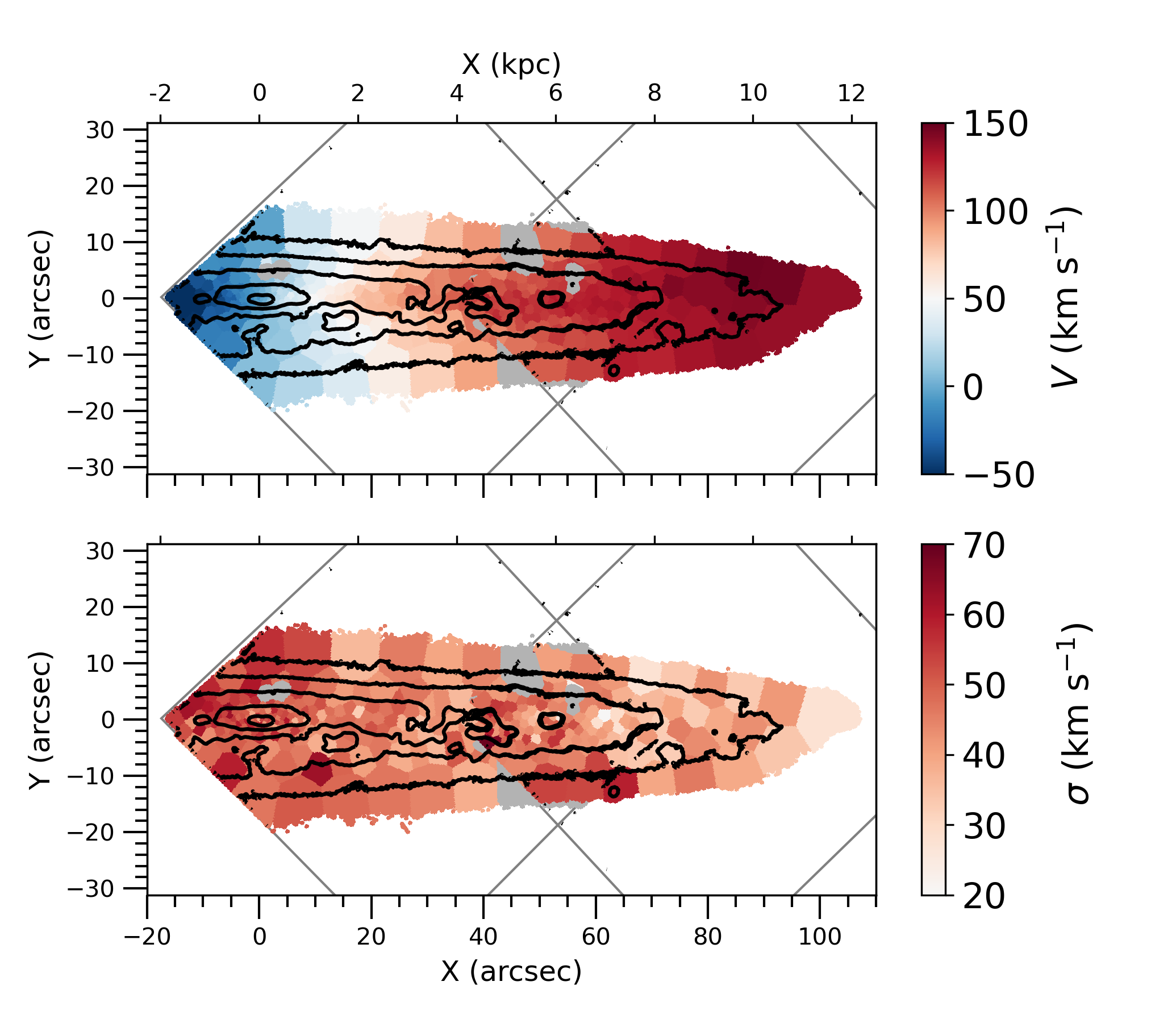}
	\end{minipage}
	\hfill{}
	\begin{minipage}[r]{\columnwidth}
		\centering
		\includegraphics[width=\textwidth]{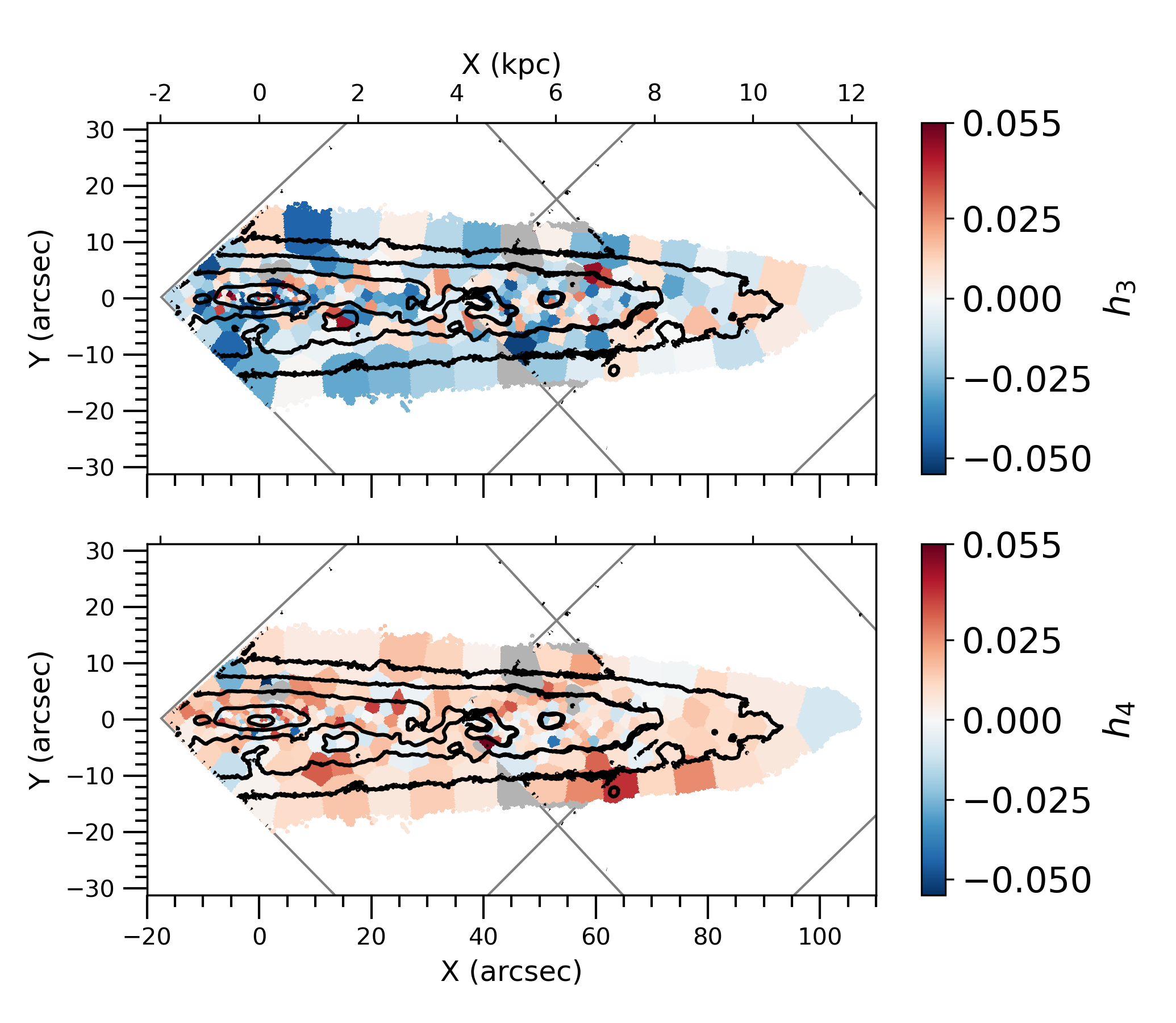}
		\label{fig:h3h4map}
	\end{minipage}
	\caption{Left: Stellar kinematic maps of the mean velocity $V$ (upper panel) and velocity dispersion $\sigma$ (bottom panel) of the stellar LOSVD. Right: Kinematic maps show values of $h_3$ and $h_4$ as results from the stellar kinematic pPXF fitting. The discarded bins are plotted in grey and the approximate positions of the MUSE pointings are indicated by grey lines, while the isophotes are displayed in black.
    }
	\label{fig:kin}
\end{figure*}

\subsection{Gas emission line fitting}
\label{subsec:gandalf}

For extracting the stellar populations, a spectrum cleaned from the gas emission lines is needed. This is important as we want to fit the absorption spectra of the galaxy without masking regions of important absorption lines like H$\beta$, the most age-sensitive feature in the analyzed range. Therefore we used the Gas and Absorption Line Fitting method (GandALF), described by \citet{sarzi_2006}, which allows us to fit both the stellar spectrum and the emission lines simultaneously, to subtract the emission lines from the rest of the spectrum and get a cleaned spectrum.
We used a customized version of the one implemented in the GIST pipeline \citep{bittner_2019} and fit the emission lines mentioned in Section~\ref{subsec:kin_fit}. The resulting maps of the ionized gas are presented and discussed in Appendix~\ref{appendix:gas_maps}.

\subsection{Stellar population fitting}
\label{subsec:pop_fitting}

We used pPXF, now with a different setup described below, to extract the stellar populations of NGC~3501. 
The combination of SSP models used by pPXF, to fit the observed spectra, allows us to extract mass- and light-weighted stellar-population parameters. For the mass-weighted results, the SSP models were kept as they are, since they are already normalized to have the mass of 1~M$_{\odot}$, but to obtain light-weighted results the SSP models were normalized by their own mean flux.

Because this galaxy has a lot of gas and dust, we first fitted each Voronoi bin spectrum with pPXF for dust extinction. 
Then we corrected the spectra for that using the output from the pPXF extinction fit together with a \citet{calzetti_2001} extinction curve. 
Additionally, we corrected all spectra for the extinction of the Milky Way following \citet{emsellem_2022}, where they used a \citet{cardelli_1989} extinction law and E(B-V) values from \citet{schlafly_2011} to determine the stellar continuum extinction.
After that, we fitted each spectrum for the stellar population parameters with pPXF again, using a multiplicative polynomial of 8$^{th}$ degree and no additive polynomials. 
We set up a regularization for the stellar population weights as suggested in \citet{cappellari_2017}. Here we used the calibration method explained e.g. in \citet{boecker_2019}, to find the maximum allowed regularization parameter consistent with the data.
With that method, we found that a wide range of values could be applied to the data.
We then chose our regularization, in that range but much lower than the maximum value obtained from the calibration, following the approach described in \citet{pinna_2019a}. 
We applied several values of regularization and examined the resulting fits. Moreover, we plotted the star formation history with different regularizations, to see if any features of age, metallicity and [Mg/Fe]-abundance will be lost with a higher value.
In the end, a value of 10 was chosen, as a good compromise between smoothing the results but still being able to identify clearly different features in the age, metallicity and [Mg/Fe] distributions.

After examining the final fits, we discarded 9 more bins because of ill GandALF or stellar population fitting. In the end, we are left with a total number of 305 Voronoi bins in both pointings that provided good fits in each step.

\subsection{Uncertainties in stellar kinematic and population fitting}
\label{subsec:uncertainties}

We performed Monte Carlo simulations to estimate the uncertainties. Therefore, we first fitted the spectra of each Voronoi bin with pPXF using the setup described in Section~\ref{subsec:pop_fitting} and calculated the wavelength-dependent residuals as the difference between the best-fit spectrum from pPXF and the initial galaxy spectrum. We then added noise to the initial spectra, where for each spectral pixel the noise is taken from a Gaussian distribution of standard deviation equal to the residuals. 
We obtained the uncertainties in all kinematic and stellar population analysis by computing the standard deviation of a thousand realizations of the input spectra. 
Typical values for the uncertainties of $V$, $\sigma$, $h_3$ and $h_4$ respectively are: $\sim$~10~km~s$^{-1}$, 20~km~s$^{-1}$, 0.02, 0.01.
The uncertainties for the light-weighted (mass-weighted) results for age, metallicity [M/H] and [Mg/Fe] respectively are: $\sim$ 1 (2)~Gyr, 0.1 (0.15)~dex, 0.05 (0.1)~dex. 
For more details, the uncertainty maps of the stellar kinematics and stellar population parameters are shown in Figure~\ref{fig:MC_kin} and \ref{fig:MC_pop}.


\section{Results}
\label{sec:res}

We start this section by describing stellar kinematics and population maps. The two single pointings were aligned using the isophotes, which are plotted in black. All the discarded bins are plotted in grey and the approximate positions of the MUSE pointings are indicated by grey lines.

\subsection{Stellar kinematics}
\label{subsec:kin_res}

The stellar kinematic maps of the mean velocity $V$, velocity dispersion $\sigma$, $h_3$ (skewness) and $h_4$ (kurtosis) are presented in Figure~\ref{fig:kin}. 

In the velocity map (left upper panel in Figure~\ref{fig:kin}), a smooth rise of the velocity going from the center outwards of the galaxy, with a maximum rotation velocity of $\sim$~150~km~s$^{-1}$ can be seen. Within the uncertainties, this maximum rotation velocity is in good agreement with the circular velocity from \citet{sheth_2010} seen in Table \ref{tab:prop}. Lines with the same velocity are vertical in the center of the galaxy and get more V-shaped in the outer parts indicating a faster rotation of the midplane in comparison to regions at larger heights. 
In the map of the velocity dispersion (left bottom panel in Figure~\ref{fig:kin}), we clearly see a radial gradient, from higher values in the center to lower values in the outer regions, with a slight rise at around 40~arcsec. A vertical gradient could not be observed. 

The map of the Gauss-Hermite parameter $h_3$ (top right map in Figure~\ref{fig:kin}), which describes how much the LOSVD deviates from a symmetrical Gaussian \citep{van-der-marel_1993, gerhard_1993}, shows overall negative values of $h_3$.
Some positive values of $h_3$ can be seen in the midplane at the center of the galaxy.
An anti-correlation with the mean velocity is usually associated with different rotating components of the galaxy \citep[e.g.,][]{krajnovic_2008, pinna_2019b}. However, because higher moments are penalized by pPXF when the SNR is low \citep{cappellari_2004}, we find very low values for $h_3$ and no components like a thin disc can be clearly identified here.
In the map of the Gauss-Hermite parameter $h_4$ (right bottom map in Figure~\ref{fig:kin}), characterizing the heaviness of the LOSVD tails compared to pure Gaussian tails
\citep{van-der-marel_1993, gerhard_1993}, one can see mostly positive values (broader profiles than a Gaussian) and some negative values across the midplane (narrower profiles than a Gaussian). With that, no structures can be clearly identified here either.

\subsection{Stellar population}
\label{subsec:pop_res}

\begin{figure*}
	\begin{minipage}[l]{\columnwidth}
		\centering
		\includegraphics[width=\textwidth]{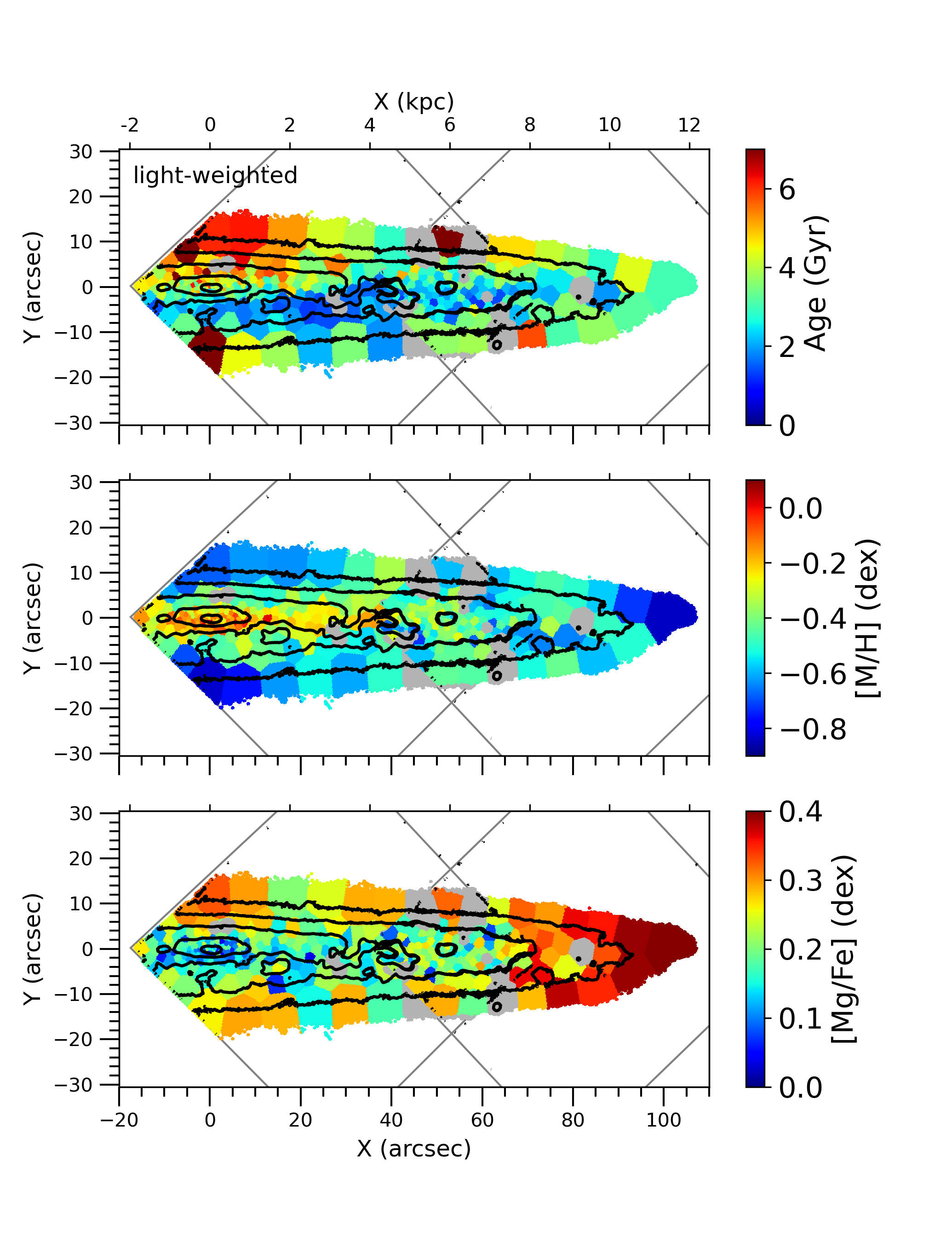}
	\end{minipage}
	\hfill{}
	\begin{minipage}[r]{\columnwidth}
		\centering
		\includegraphics[width=\textwidth]{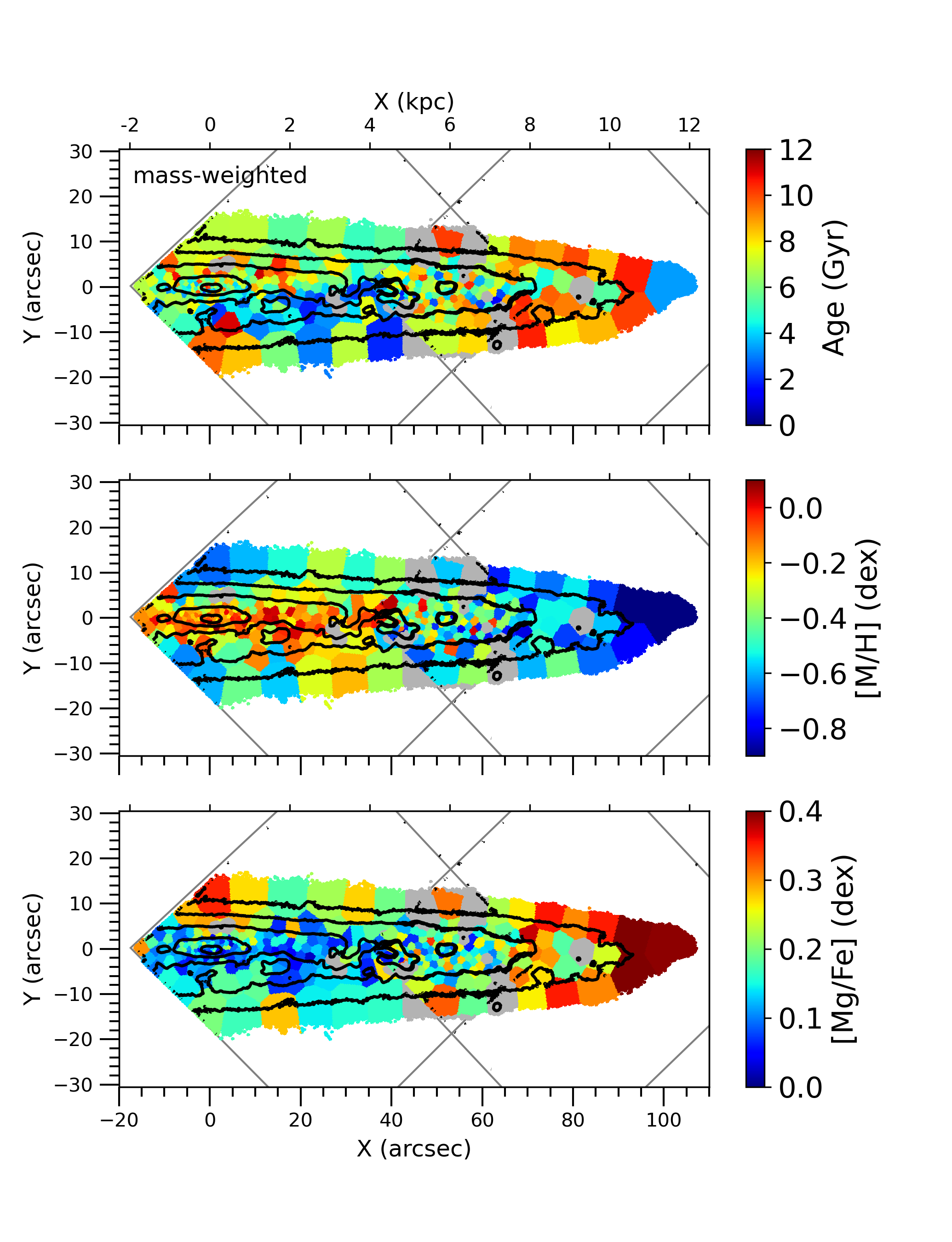}
	\end{minipage}
	\caption{Left: Stellar population maps showing the light-weighted age (top), total metallicity [M/H] (middle) and [Mg/Fe]-abundance (bottom). Right: Stellar population maps showing the mass-weighted age (top), total metallicity [M/H] (middle) and [Mg/Fe]-abundance (bottom). The approximate positions of the MUSE pointings are indicated by grey lines and the discarded bins are displayed in grey, while the isophotes are displayed in black.
    }
	\label{fig:pop}
\end{figure*}

The maps of the light-weighted (left panels) and mass-weighted (right panels) stellar populations are presented in Figure~\ref{fig:pop}. Because every SSP model matches a combination of age, metallicity [M/H] and [Mg/Fe]-abundance, the weights given to them by pPXF indicate how important each combination is for every individual bin. Therefore, each bin shows a weighted average of the stellar population parameters. In the maps, mass- and light-weighted age, metallicity [M/H] and [Mg/Fe]-abundance are displayed.

In the age maps (top maps in Figure~\ref{fig:pop}), we see a horizontal stripe of very young stars directly below the central midplane. This stripe extends for 0~$<$~X~$<$~40~arcsec in the mass-weighed map and up to 70 arcsec in the light-weighted map. The weight of younger stars is enhanced in light-weighted results, since they are more dominant in light compared to older stars. On the other hand, older stars are more dominant in mass, so they appear more pronounced in mass-weighted results.
This is also the reason why the light-weighted map shows on average younger stars than the mass-weighted map in the whole galaxy (also seen in e.g. \citealt{sanchez-blazquez_2014}).
As seen in Figure~\ref{fig:sdss_s4g}, dust is visible close to the midplane, blocking some of the light from stars in this region. So here the light of stars which are closer in the line-of-sight and tend to be younger is also integrated.
But larger gaps in the dust, that can be seen in the SDSS image of NGC~3501 in Figure~\ref{fig:sdss_s4g}, ensure that most of the light integrated along the line-of-sight has its origin in the galaxy midplane region laying behind the dust. 
The rest of the galaxy shows a mixture of mostly intermediate-age stars and some very old stars with ages going up to 12 Gyr in the mass-weighted map.
We can identify the very young stars in the midplane of the light-weighted age map as an inner thinner disc-like structure embedded in the older, extended and thicker disc. This difference is less clearly seen in the mass-weighted map.

Looking at the metallicity [M/H] maps (middle maps in Figure~\ref{fig:pop}), the highest metallicity can be found in the inner midplane of the galaxy (with some regions reaching a solar metallicity), surrounded by lower metallicity values in the outer and thicker disc. In the midplane, some spots with solar metallicity are visible.
This negative vertical gradient is stronger in the light-weighted results than in the mass-weighted ones, but is clearly visible in both maps and also in both MUSE pointings.
The mass-weighted map also shows a negative radial gradient in metallicity. In the light-weighted metallicity map, this gradient is not so pronounced.
Comparing the light-weighted and mass-weighted metallicities, the stars in the mass-weighted maps are more metal-rich than the ones in the light-weighted maps, especially those in the midplane and the center of the galaxy.
Furthermore, the mass-weighted metallicity map had overall higher values than the light-weighted map and with Monte Carlo simulations, we obtained on average slightly larger uncertainties for mass-weighted ages, metallicities and [Mg/Fe]-abundances. For more details, maps of the uncertainties are shown in Figure~\ref{fig:MC_pop}.

The maps of the [Mg/Fe]-abundance (bottom maps in Figure~\ref{fig:pop}) show a clear anti-correlation with the metallicity maps. The highest super-solar [Mg/Fe] values are located in the outer regions. The lowest [Mg/Fe]-abundance can be found in the center of the galaxy, best visible as a small blueish stripe in the center of the light-weighted map. In the mass-weighted one, this region with low [Mg/Fe] values extends further to the outer regions, but is still more dominant in the central midplane, reaching partially solar abundances. Positive vertical and radial gradients are visible in both maps, whereas the radial gradient is more pronounced in the mass-weighted abundances.
The inner and thinner disc-like structure of very young stars in the midplane appears to be also the most metal-rich and least [Mg/Fe] enhanced, embedded in the older, metal-poor and [Mg/Fe]-enhanced thicker disc structure.

\subsection{Star formation history}
\label{subsec:sfh_res}

\begin{figure}
	\includegraphics[width=\columnwidth]{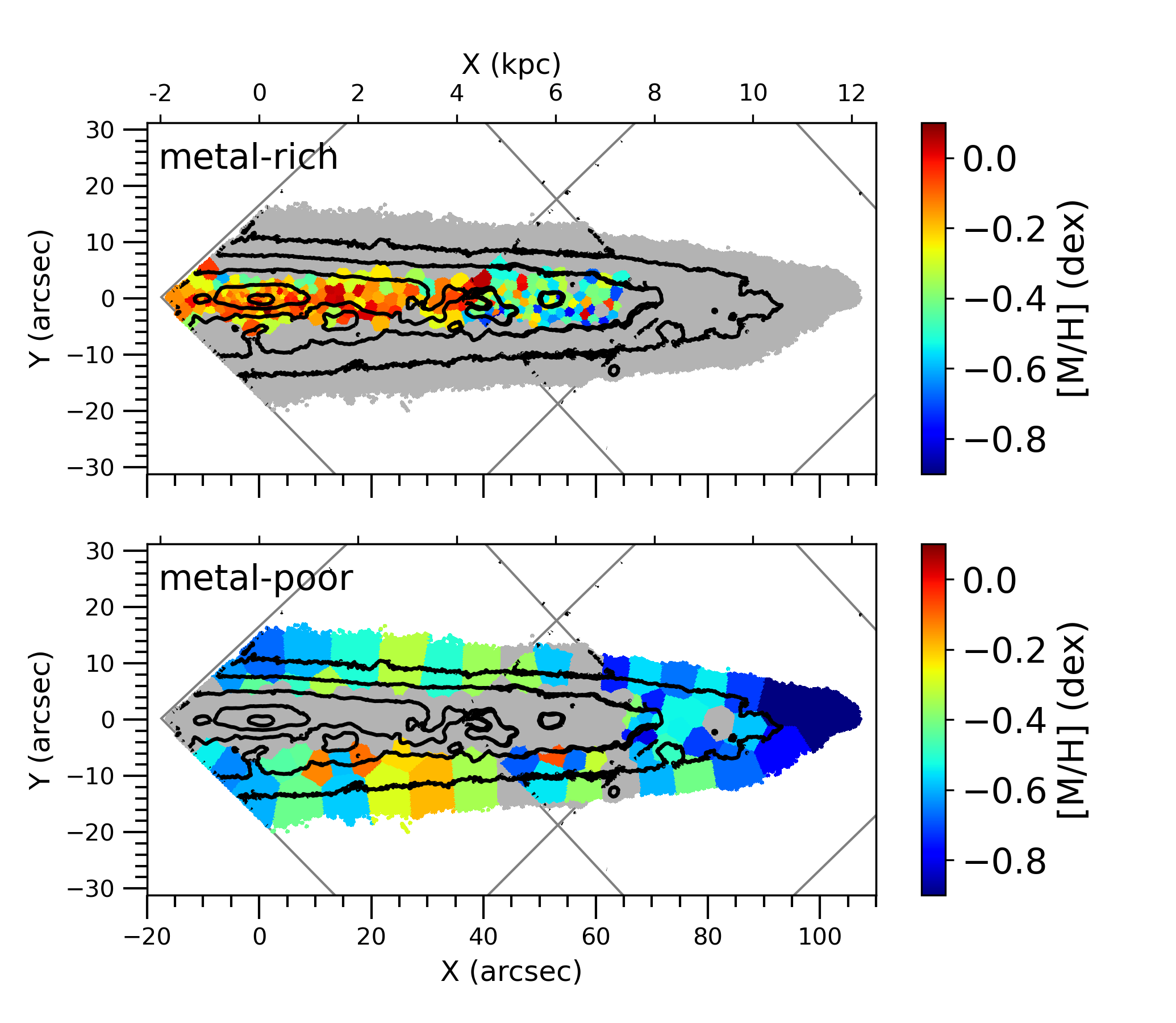} 
	\caption[]{Mass-weighted metallicity maps, where the metal-rich (top) and metal-poor (bottom) components are indicated. In each panel, all the bins of the other component are plotted in grey and the approximate positions of the MUSE pointings are indicated by grey lines, while the isophotes are displayed in black.
    }
	\label{fig:comp}
\end{figure}

Because of the differences in age, metallicity [M/H] and [Mg/Fe] between different regions of the galaxy, described in Section~\ref{subsec:pop_res}, we define two components: metal-rich and metal-poor.
This separation is based on the large jumps in the vertical and radial metallicity and [Mg/Fe] abundance. The highest metallicity and lowest [Mg/Fe] bins are therefore included in the metal-rich component, while the lower metallicity and higher [Mg/Fe] bins are included in the metal-poor component.
Those components are visualized in Figure~\ref{fig:comp}, using the mass-weighted metallicity map (Figure~\ref{fig:pop}, right).

\begin{figure*}
	\begin{minipage}[l]{\columnwidth}
		\centering
		\includegraphics[width=\textwidth]{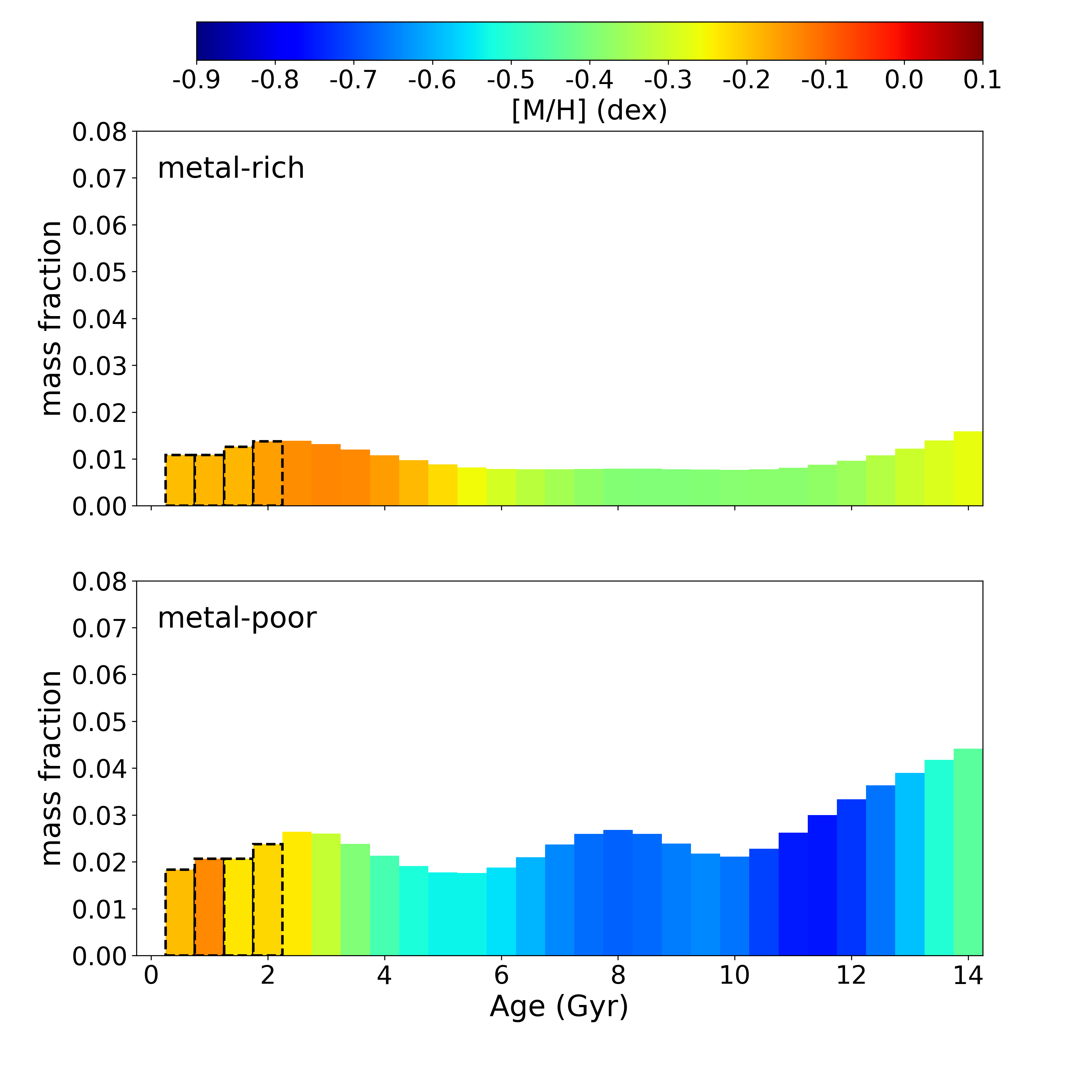}
	\end{minipage}
	\hfill{}
	\begin{minipage}[r]{\columnwidth}
		\centering
    	\includegraphics[width=\textwidth]{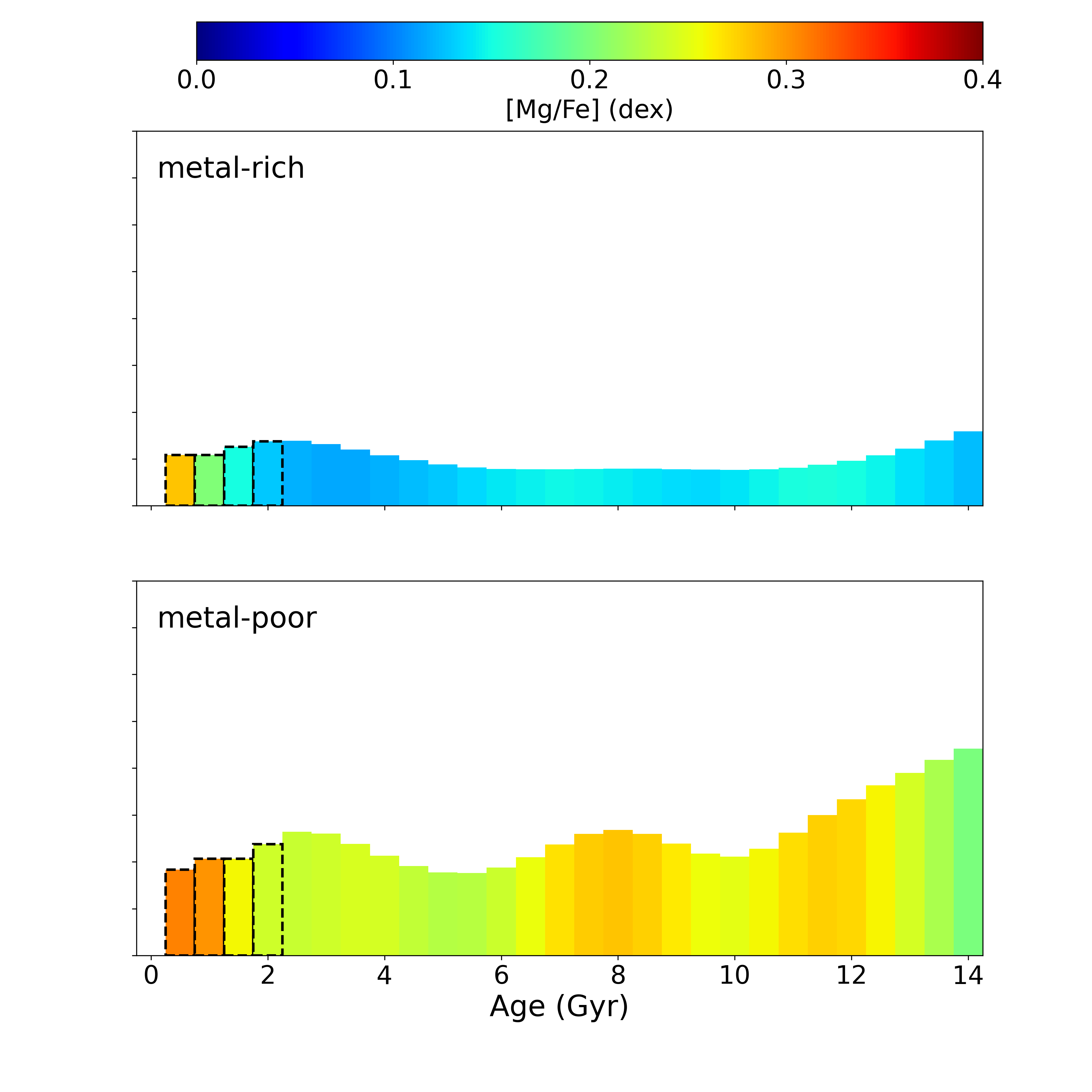}
	\end{minipage}
	\caption{Star formation history in terms of the mean mass fraction (normalized to the galaxy mass in the region covered by our data) in the metal-poor (bottom) and metal-rich (top) components as a function of age. Age bins are color-coded by metallicity [M/H] (left), and by [Mg/Fe] (right). Black dashed edges mark age bins with [Mg/Fe] abundances that we do not consider totally reliable, because of difficulties in retrieving the [$\alpha$/Fe] for young stars as discussed in Appendix~\ref{appendix:system_unc}.
    }
	\label{fig:metalpha}
\end{figure*}

To get the star formation history of each component normalized by the total stellar mass of the covered galaxy half, we calculated the mass of each Voronoi bin as described in \cite{pinna_2019a}. For that, we used an SDSS g-band image of NGC~3501 from DR16 \citep{ahumada_2020}, as well as the mass-to-light ratios in each Voronoi bin calculated by combining our mass-weighted results with the corresponding mass-to-light ratios to our models from \cite{vazdekis_2015}. For the region that is covered by our Voronoi bins, we derived a total stellar mass of 5~$\times$~10$^9$~M$_{\odot}$. This value is compatible with the total stellar mass in Table~\ref{tab:prop}, calculated in the 3.6 and 4.5~$\mu$m bands, by also taking into account that half of the galaxy mass is missing in our two MUSE pointings and a significant amount of stellar light in the optical range may be obscured by the dust.
Moreover, the metal-rich component covers $28\%$ of the Voronoi binned region, making up $1.4$~$\times$~$10^9$~M$_{\odot}$, whereas the metal-poor component covers $72\%$, making up $3.6$~$\times$~10$^9$~M$_{\odot}$.

The resulting star formation history is plotted in Figure~\ref{fig:metalpha}, in terms of the mass fraction of the covered galaxy as a function of age. The age bins are color-coded by the weighted-average metallicity [M/H] (left) and the weighted-average [Mg/Fe] (right). Those values were derived as the average metallicity [M/H] or [Mg/Fe] for each age bin in the component, weighted by their respective mass fractions. The star formation history of the metal-rich component is plotted at the top, whereas the star formation history of the metal-poor component is plotted at the bottom.

Looking only at the slope of the star formation history, the metal-rich and metal-poor components peak for old ages around 14 Gyr. Later on, the star formation slightly decreases but does not stagnate and peaks again at youngest ages in both components. So the metal-rich component forms steadily. For the metal-poor component, we also see a peak around 8 Gyr, so different bursts of star formation are visible here.

Uncertainties for the mass fractions indicated in these star formation histories are $\sim$ 0.003 (0.015) for the metal-rich (metal-poor) component. The detailed uncertainties, computed from Monte-Carlo simulations, are shown in Appendix~\ref{appendix:uncertainties}.

\subsection{Chemical evolution}
\label{subsec:chem_comp}

If we look at the color-coded bars (Figure~\ref{fig:metalpha}), we see for both components sub-solar metallicities at all ages. 
On average, the metal-rich component has by definition an overall higher metallicity (from $-$0.1~dex to $-$0.4~dex) and an overall lower [Mg/Fe] than the metal-poor component (from 0.1~dex to 0.2~dex). By contrast, the metal-poor component has a more extended metallicity range (from around $-$0.1~dex to $-$0.8~dex) and its [Mg/Fe] reaches higher values (from around 0.2~dex to 0.3~dex).

Starting with the metal-rich component (top), we see that stars formed in the peak at 14~Gyr do not show the lowest metallicity nor the highest [Mg/Fe]. The metallicity then slightly decreases and stays relatively steady till $\sim$~5~Gyr. After that, the metallicity increases to nearly solar values, while the star formation history reaches its second big peak at the latest ages. During that, the [Mg/Fe] oscillates slightly until 0.5~Gyr, where it increases to higher abundances. 
Continuing with the metal-poor component (bottom), we see at the oldest ages a higher mass fraction with a lower metallicity and higher [Mg/Fe] compared to the metal-rich component. Younger populations at around 12~Gyr then have a significantly lower metallicity. From 12~Gyr on, towards the youngest ages, we observe oscillations in metallicity, following a chemical enrichment that alternates with metallicity drops. In this metal-poor component, [Mg/Fe] also oscillates as anti-correlated to metallicity, where higher [Mg/Fe]-abundances correspond to peaks in star formation as well as to drops in metallicity. The highest [Mg/Fe]-peak then appears at the youngest ages.

So while the inner midplane evolves rather slowly and evenly with a more balanced [Mg/Fe] ratio, the more extended and thicker metal-poor component suffered from several bursts leading to the formation of more metal-poor and [Mg/Fe]-enhanced stars.
Uncertainties for the metallicity [M/H] and [Mg/Fe] values indicated in these star formation histories are for the metal-rich (metal-poor) component: $\sim$~0.15 (0.3)~dex in [M/H] and $\sim$~0.05 (0.1)~dex in [Mg/Fe]. The detailed uncertainties, computed from Monte-Carlo simulations, are shown in Appendix~\ref{appendix:uncertainties}. 
The youngest bins with high [Mg/Fe] abundance are marked with dashed edges to warn the reader about the low reliability of those values. This and other aspects are extensively discussed in Appendix~\ref{appendix:system_unc}.


\section{Discussion}
\label{sec:dis}

In spite of the potential large systematic uncertainties for the youngest stars, and being aware of the limitations of our method (all discussed in Appendix~\ref{appendix:system_unc}), we strongly support the oscillations in [Mg/Fe] and metallicity for older ages in the star formation histories (Figure~\ref{fig:metalpha}). These are discussed in the following sections.

\subsection{Disc structure}
\label{subsec:disc}

Our kinematic maps (Figure~\ref{fig:kin}) do not show two clearly distinct disc structures, in agreement with the morphological decomposition from \citet{comeron_2018}, who suggested that NGC~3501 consists of a single thicker stellar disc. 
A sharp vertical jump in velocity dispersion would imply a two-disc structure for the galaxy, as it is observed in other edge-on galaxies \citep[e.g.,][]{pinna_2019a, pinna_2019b} that have distinct thick and thin discs, where the velocity dispersion is higher in the thick disc than in the thin disc.
The rise in velocity dispersion at around 40~arcsec might be the result of turbulent star formation in a spiral arm structure of the galaxy. This star formation region is also visible in Figure~\ref{fig:gas_flux}, where one can see the high flux of H$\beta$ and [OIII] emission lines in this specific region of higher stellar velocity dispersion, which also has a higher velocity dispersion in the gas component, as it can be seen in Figure~\ref{fig:gan}.

Because the galaxy is not seen completely edge-on, some part of the boxy bulge is visible at the top center of the galaxy with a higher velocity dispersion. However, the bulge is faint and we do not see a box-shaped pattern of high values in velocity dispersion in the bulge region (as it was found e.g. in \citealt{pinna_2019a}). So the box-shape from \cite{luetticke_2000} could not be observed in our kinematic maps. We cannot provide more details on the bulge since it cannot be identified either in any of the stellar population maps.

We know that most galaxies displaying a thin and thick disc show concrete differences in their vertical age distribution, where the thin disc shows an overall younger population than the thick disc \citep[e.g.,][]{yoachim_2008b, kasparova_2016, scott_2021, martig_2021}. 
In contrast to early-type galaxy observations, we see ongoing star formation in NGC~3501, so stellar populations of this galaxy are rather young and uniform (Figure~\ref{fig:pop}). This is also in good agreement with the ages derived in \citet{marino_2010} (see Section~\ref{sec:gal}) and their findings that NGC~3501 is in an unvirialized phase, which could be the reason for active star formation.

The metallicity distribution of mass- and light-weighted results (Figure~\ref{fig:pop}) shows negative gradients from average solar and slightly sub-solar metallicities for the stars in the inner midplane to sub-solar metallicities and lowest values in the outer envelope of the galaxy. While the [Mg/Fe] distribution showed positive gradients from nearly solar values in the midplane to super-solar values in the outer galaxy parts. 

The negative radial metallicity gradient is also seen by \citet{marquez_2002} for this galaxy (see Section~\ref{sec:gal}).
Sharp vertical gradients can be seen in \citet{comeron_2016, guerou_2016} (only metallicity) and \citet{pinna_2019a, pinna_2019b} (metallicity and [Mg/Fe]) for their analyzed S0 galaxies, where high super-solar metallicities and nearly solar [Mg/Fe] abundances are found in the midplane and lower metallicities and super-solar [Mg/Fe] in the thick disc. The differences in metallicity are more prominent in \citet{pinna_2019a, pinna_2019b} and NGC~3501 than in \citet{comeron_2016, guerou_2016}.
The overall anti-correlation of [Mg/Fe] to metallicity that we see in the maps and the star formation history was observed too in other external galaxies and in the Milky Way  \citep[e.g.,][]{pinna_2019a, pinna_2019b, bittner_2020, wheeler_2020}. 
Although S0 and Sc galaxies have different morphologies and had different evolution histories, the differences in S0 metallicities associated with the presence of a thick and a thin disc, help us to associate the differences in our metallicity maps to different disc components, with probably different origins.
Also, the Milky Way shows negative metallicity and positive [$\alpha$/Fe] gradients, where more solar metallicities and [$\alpha$/Fe] abundances are located in the midplane and sub-solar metallicities and slightly super-solar [$\alpha$/Fe] values in the thicker disc at larger distances from the midplane \citep{minchev_2019, wheeler_2020, gaia-collaboration_2022}, while having some flaring of the thin disc at larger radii. 
This metallicity and [$\alpha$/Fe] gradients in the Milky Way were associated with an ''inside-out'' scenario, where the more metal-rich and $\alpha$-poor thin disc formed from the inside-out embedded in a more metal-poor and $\alpha$-rich thick disc \citep[e.g.,][]{minchev_2013, schoenrich_2017, frankel_2019, gaia-collaboration_2022}.
Moreover, \citet{scott_2021} found $\alpha$-rich and $\alpha$-poor stellar populations in the Milky Way-like galaxy UGC 10738, that have a similar spatial distribution to the Milky Ways $\alpha$-rich and $\alpha$-poor populations.

One can also see from Figure~\ref{fig:comp} that our metal-rich component, which extends to around 65~arcsec, agrees very well with the first break radius from \citet{martin-navarro_2012} at $\sim$~67~arcsec (see Section~\ref{sec:gal}), related to a threshold in ongoing star formation. This threshold can be seen in the distribution of younger stars in the metal-rich component, whereas the metal-poor component consists of older stars on average. The truncation radius at $\sim$~100~arcsec \citep{martin-navarro_2012}, which is rather a real drop in the stellar mass density of the disc, lies near the limit of our Voronoi binning at around 108~arcsec. Here, the Voronoi binning ends because of the given SNR-threshold for the spaxels. The SNR is lower for lower luminosity regions, caused by the lower stellar mass density in the outer regions of the galaxy.

So stars with younger ages, higher metallicities and lower [Mg/Fe]-abundances in the inner midplane, embedded within older stars, with sub-solar metallicities and super-solar [Mg/Fe]-abundances, suggest that two disc components are present. Moreover, in the mid-infrared image of NGC~3501 from \citet{comeron_2018}, less affected by the dust in the galaxy than our MUSE data in the optical range, a thin and much brighter structure is visible in the inner midplane. However, \citet{comeron_2018}, aiming at fitting vertical profiles in different radial bins to identify different disc components, excluded this central region which extends between $-$25 and 25~arcsec in the radial direction, because it may be contaminated by the bulge. 
That brighter inner region spatially matches the young age, high-metallicity and low-[Mg/Fe]-abundance structure in the central midplane, seen on our stellar population maps (Figure~\ref{fig:pop}). So we propose the presence of two structural disc components in that radial range: a thicker metal-poor and older main component, with a much less radially extended metal-rich thin disc, recently formed.

\subsection{Formation and evolution of NGC 3501}
\label{subsec:formation}

In the following, we will place the results of the stellar kinematics, population and star formation history analysis of NGC~3501 in context with its formation history.

In the star formation history (Figure~\ref{fig:metalpha}) of the metal-poor component (bottom panels), we see higher metallicities at very old ages with a significant drop later around 12~Gyr ago. This drop is also seen in the metal-rich component, but less enhanced. 
Such relatively high metallicities at early times are also seen in closed-box chemical evolution models by \citet{vazdekis_1996}, where high metallicities can be achieved very quickly with very intense star formation.
Another possibility to produce higher metallicities at old ages is to accrete metal-rich old stars during the evolution process of the galaxy \citep{feuillet_2022}. 

To explain the drop in metallicity around 12~Gyr ago, associated with an increase of [Mg/Fe], we propose the accretion of a large mass of pristine gas from the outside, followed by a burst of star formation from that external gas. 
A second burst of star formation from more pristine gas explains as well the more metal-poor and [Mg/Fe]-enhanced peak in the star formation history of the metal-poor component about 8~Gyr ago.  
So the bursts in the metal-poor component correspond to different episodes of gas accretion. 
Our interpretation of these oscillations in metallicity and [Mg/Fe] is the combination of the internal chemical enrichment, similar to what we would have in a closed-box, and the contribution of bursts of star formation from external material. 
The metal-rich component is not very affected by those bursts and seems to keep forming stars at the same pace. So in general, the external parts of the galaxy might be more affected by accretion.

\citet{marquez_2002} suggested that NGC~3501 and the nearby face-on galaxy NGC~3507 may be mildly interacting, using the region of H$\alpha$ emission. No signs of this mild interaction were found in the optical using SDSS by \citet{knapen_2014} though. These interactions between NGC~3501 and other galaxies in LGG~225 could drive the pristine gas into NGC~3501 and enhance the star formation events. The drop in metallicity could then come from metal-poor gas inflow along filaments or even from the interactions with surrounding galaxies. 

The youngest starburst is observed to be the most metal-rich and $\alpha$-enhanced.
This starburst might also explain the higher [Mg/Fe] values in the light-weighted maps, compared to the mass-weighted maps, since the former are more biased towards younger stars than the latter (see also Section~\ref{subsec:pop_res}).
Usually, [Mg/Fe] increases with age \citep[e.g.,][]{rahimi_2011, nissen_2015} and would rather be lower than higher for younger stars. 
Also, most of the results from \citet{bittner_2020} show equal [$\alpha$/Fe]-abundances for mass- and light-weighted results. Only a few galaxies from their sample show lower [$\alpha$/Fe]-abundances in the mass-weighted maps, which the authors explained with higher uncertainties in their mass-weighted [$\alpha$/Fe] maps. 
This young population of higher [Mg/Fe] and metallicity here, could rather be the result of a sudden and significant increase in the star formation efficiency \citep{conroy_2022}.

As seen in Section~\ref{subsec:sfh_res} for NGC~3501, the stellar mass of the thicker metal-poor component is currently much larger than the one of the thinner metal-rich component.
\citet{comeron_2011} show that the thin disc typically is the more dominant stellar mass component for galaxies with a circular velocity of around 150~km~s$^{-1}$.
However, a comparison with this trend from \citet{comeron_2011} 
is complicated, since NGC~3501 is atypical. The thinner metal-rich component of NGC~3501 has not fully formed yet and the metal-rich and metal-poor components of NGC~3501 are not comparable to typical thin and thick discs (e.g. in scale height, velocity dispersion, mass).
Also, if we extrapolate the trends of the star formation history in Figure~\ref{fig:metalpha}, the thicker metal-poor component seems to keep forming stars at a similar pace as the thinner metal-rich component.
Since this region of higher metallicity will likely grow into the parts of the galaxy that are more metal-poor now and we currently do not know how far the metal-rich region will expand until it becomes a more typical thin disc, it is difficult to measure the star formation rate belonging to a future thin disc and give a secure estimate on the mass evolution of the metal-rich component right now.

It is important to mention that the thicker, metal-poor disc is also present in the inner midplane, overlapping in the line-of-sight to the metal-rich component. This could bias the inner metal-rich region towards lower metallicities, higher [Mg/Fe] and older stars on average, contributing some features of the metal-poor component (lower metallicity in intermediate ages, higher [Mg/Fe] around 8 and 12 Gyr, higher fraction of stellar mass around 14 Gyr in age) in the star formation history of the metal-rich one.
We checked for contamination of the thicker metal-poor disc in the metal-rich component in radial direction by restricting the metal-rich component only to around 40~arcsec. We found that the continuous star formation in the metal-rich component, as well as values for metallicity and [Mg/Fe] do not change in a relevant amount. 

Altogether, older ages, low metallicities and the [Mg/Fe] enhancement suggest that the thicker more-extended region of the galaxy formed faster. By contrast, the younger metal-rich inner midplane in NGC~3501 kept evolving from sub-solar to solar metallicities and a more balanced [Mg/Fe]-abundance. 
In terms of galaxy evolution, \citet{yu_2021} showed that each galaxy has a bursty period of star formation in early times, leading into a steady phase having a nearly constant star formation rate. Stars in the bursty phase are mostly scattered into hot orbits, not being arranged in a thin disc yet. This bursty phase then ends and the steady phase begins, where a thin disc becomes more pronounced. Similarly, in the ''born-hot'' scenario \citep{brook_2004}, the thick disc forms in-situ at high redshift and the thin disc forms from gas accretion during multiple mergers. The stars forming in the bursty phase are today found in a thick disc structure, whereas young stars in the steady phase are located in a thin disc. This scenario could explain the more bursty star formation history of NGC~3501s metal-poor component, while also leading to a more steady star formation history of the metal-rich component.
A positive radial [Mg/Fe] gradient might be a feature of galaxies formed through an ''outside-in'' formation scenario, because the outer disc develops earlier and faster than the inner region, probably leaving outer radii in higher [Mg/Fe]-abundances (shown for elliptical galaxies in \citealt{pipino_2004, pipino_2006} and low-mass galaxies in e.g., \citealt{gallart_2008, zhang_2012}). 
NGC~3501 also shows positive vertical and radial [Mg/Fe] gradients, supporting this scenario. Also, the star formation history suggests that the extended component was already born as a thicker disc during a bursty period, and was undergoing several bursts of star formation. By contrast, the inner midplane evolved more steadily and might be less affected by incoming material from the outside than the outer envelope. 
While the full galaxy may be following an ''outside-in'' formation since the outer metal-poor component was formed first, and only later the inner metal-rich disc, the latter may be the early innermost region of a future extended thinner disc, forming ''inside-out''.\\


\section{Conclusion}
\label{sec:con}

Two-dimensional maps of the stellar kinematics and populations as well as the star formation history of the late-type galaxy NGC~3501 have been obtained, using deep MUSE integral-field spectroscopy. Two MUSE pointings were used to observe the south-western half of the galaxy, which is seen edge-on. Maps of the mean velocity and velocity dispersion support a single disc-like rotation with a slightly faster rotation of the midplane region. We were able to see higher velocity dispersion at the center of the galaxy, which may coincide with the central boxy bulge of the galaxy \citep{luetticke_2000}. 
Maps of the mass- and light-weighted stellar age, metallicity [M/H] and [Mg/Fe]-abundance were presented. These maps showed that the galaxy consists mostly of young stars with a positive vertical and radial gradient in age, clearer in the light-weighted maps. The metallicity maps show solar values in the inner midplane, with a vertical and radial gradient to sub-solar metallicities in the outer envelope.
The [Mg/Fe] abundance showed solar values in the inner midplane and super-solar abundances at larger radii and heights. Those gradients and the different star formation histories indicate a small thin disc existent in NGC~3501. 

During the first stages of the evolution of NGC~3501, there was a single thicker disc probably born already thick in a bursty phase, with global lower metallicity and more significant [Mg/Fe] enhancement than components formed later. 
On the other hand, in the inner midplane, a thin disc-like structure slowly developed in a more steady phase, leading to solar metallicities and a lower [Mg/Fe]-abundance in the later phases of its evolution.
Since NGC~3501 is gas-rich and mildly interacting, external events might have brought more-pristine gas with lower metallicities and more enhanced in $\alpha$-elements, leaving several more [Mg/Fe]-enhanced populations formed from this external material in the outer parts of the galaxy.
Because of its young age, NGC~3501 might be a galaxy in an early stage of galaxy evolution, showing a snapshot of the early evolutionary phase of a thin disc embedded in a thick disc. 
This galaxy will very likely form a more pronounced massive thin disc in the future, where the current central midplane already builds a base.\\



\section*{Acknowledgements}

This work was funded by the Deutsche Forschungsgemeinschaft (DFG, German Research Foundation) -- Project-ID 138713538 -- SFB 881 (``The Milky Way System'', subproject B08).\\
F.P. and J.~F-B acknowledge support through the RAVET project by the grant PID2019-107427GB-C32 from the Spanish Ministry of Science, Innovation and Universities (MCIU), and through the IAC project TRACES which is partially supported through the state budget and the regional budget of the Consejer\'ia de Econom\'ia, Industria, Comercio y Conocimiento of the Canary Islands Autonomous Community.\\
GvdV acknowledges funding from the European Research Council (ERC) under the European Union's Horizon 2020 research and innovation program under grant agreement No 724857 (Consolidator Grant ArcheoDyn).


\section*{Data Availability}

The data described in this article is available on the ESO Science Archive Facility in the program 098.B-0662, PI: Pinna, F.


\bibliographystyle{mnras}
\bibliography{bib}


\appendix

\section{Gas maps}
\label{appendix:gas_maps}

\begin{figure}	     
        \includegraphics[width=\columnwidth]{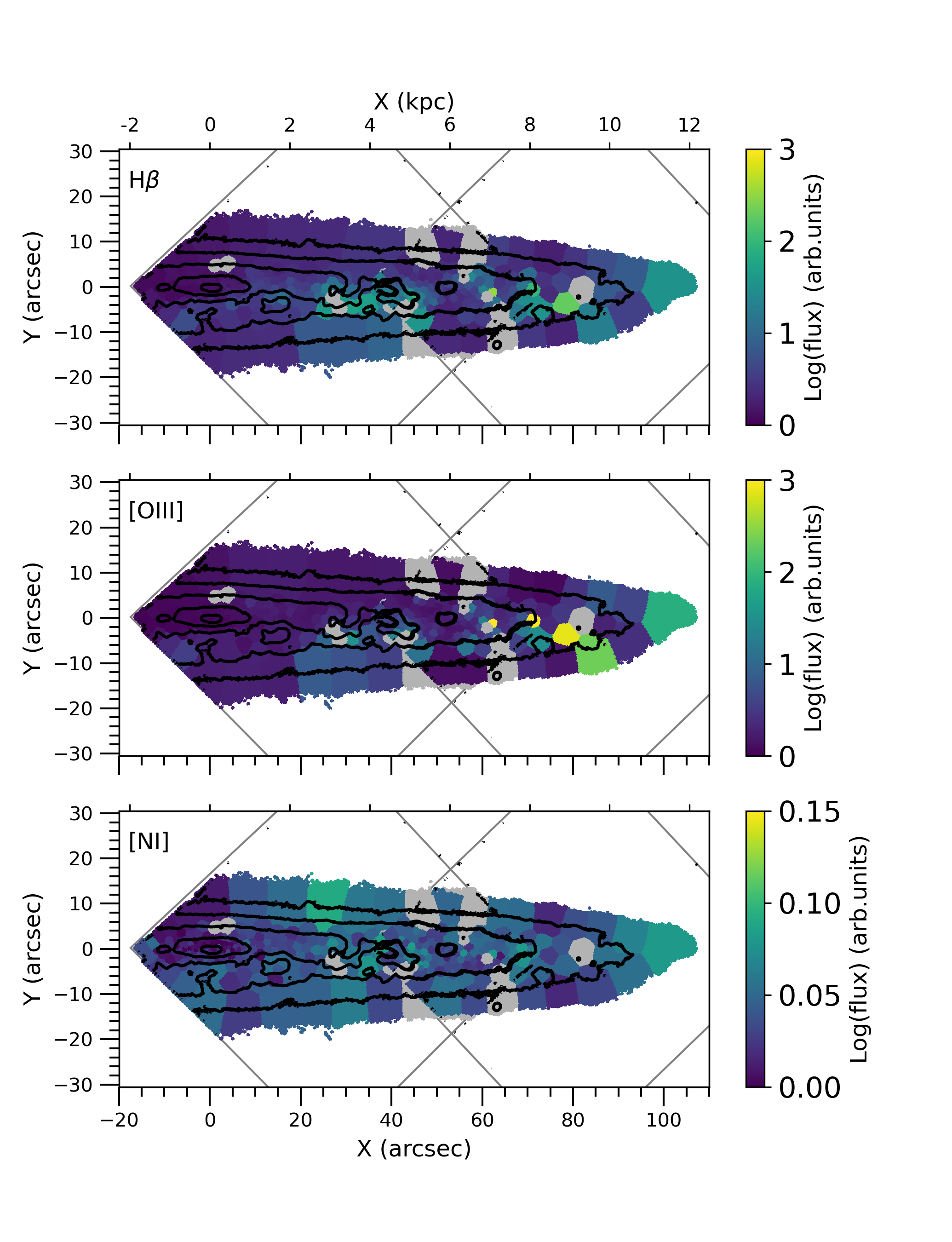} 
	\caption[]{Flux maps of the three fitted emission lines H$\beta$, [OIII], [NI] (from top to bottom).
	The discarded bins are plotted in grey and the approximate positions of the MUSE pointings are indicated by grey lines, while the isophotes are displayed in black.
    }
	\label{fig:gas_flux}
\end{figure}

\begin{figure*}
	\includegraphics[width=\textwidth]{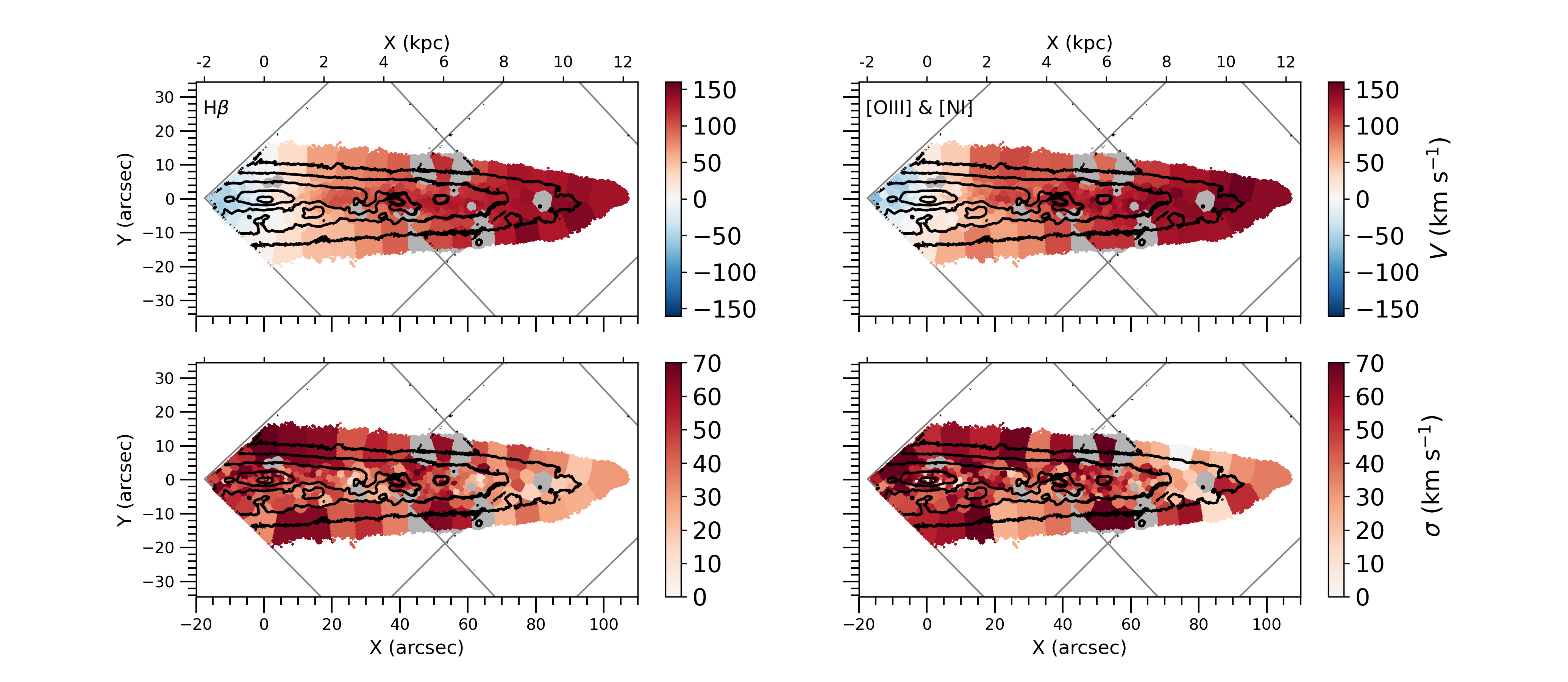} 
	\caption[]{Kinematic maps of the ionized gas determined from fitting the emission lines H$\beta$ (left), and the doublets of [OIII] and [NI] (right). The mean velocity $V$ is plotted in the top panels and the mean velocity dispersion $\sigma$ in the bottom panels. The kinematics of the [NI] emission line were fixed to the [OIII] kinematics, so the kinematics are the same. 
	The discarded bins are plotted in grey and the approximate positions of the MUSE pointings are indicated by grey lines, while the isophotes are displayed in black.
    }
	\label{fig:gan}
\end{figure*}

From the GandALF emission line fitting, the kinematics and the fluxes of the ionized gas from the three emission lines: 
H$\beta$ (4861.33~$\si{\angstrom}$), [OIII] doublet (4958.92 \& 5006.84~$\si{\angstrom}$) and [NI] doublet (5197.90 \& 5200.39~$\si{\angstrom}$) were obtained.

Figure~\ref{fig:gas_flux} shows the fluxes of the fitted emission lines, where regions of high fluxes could be the result of some spiral arm structure with a higher star formation rate.
The gas kinematic maps, presented in Figure~\ref{fig:gan} show the mean velocity $V$ (top maps) and velocity dispersion $\sigma$ (bottom maps), measured from the emission lines.
The gas kinematics of the [NI] emission line were fixed to the [OIII] kinematics for the fitting with GandALF, so the kinematics show identical results.
We can see in the gas velocity maps, that the gas is overall co-rotating with the stellar material of the galaxy disc.
Moreover, some pattern can be seen in the velocity dispersion, where regions of lower velocity dispersion correspond to regions of higher flux in H$\beta$ and [OIII] (see Figure~\ref{fig:gas_flux}), except for the region with high stellar velocity dispersion at $\sim$~40~arcsec.

Performing more detailed emission line-based measurements is beyond the scope of this paper and will be done in a following project.

\section{Uncertainties in stellar kinematics, stellar populations and the star formation history}
\label{appendix:uncertainties}

\begin{figure*}
	\begin{minipage}[l]{\columnwidth}
		\centering
		\includegraphics[width=\textwidth]{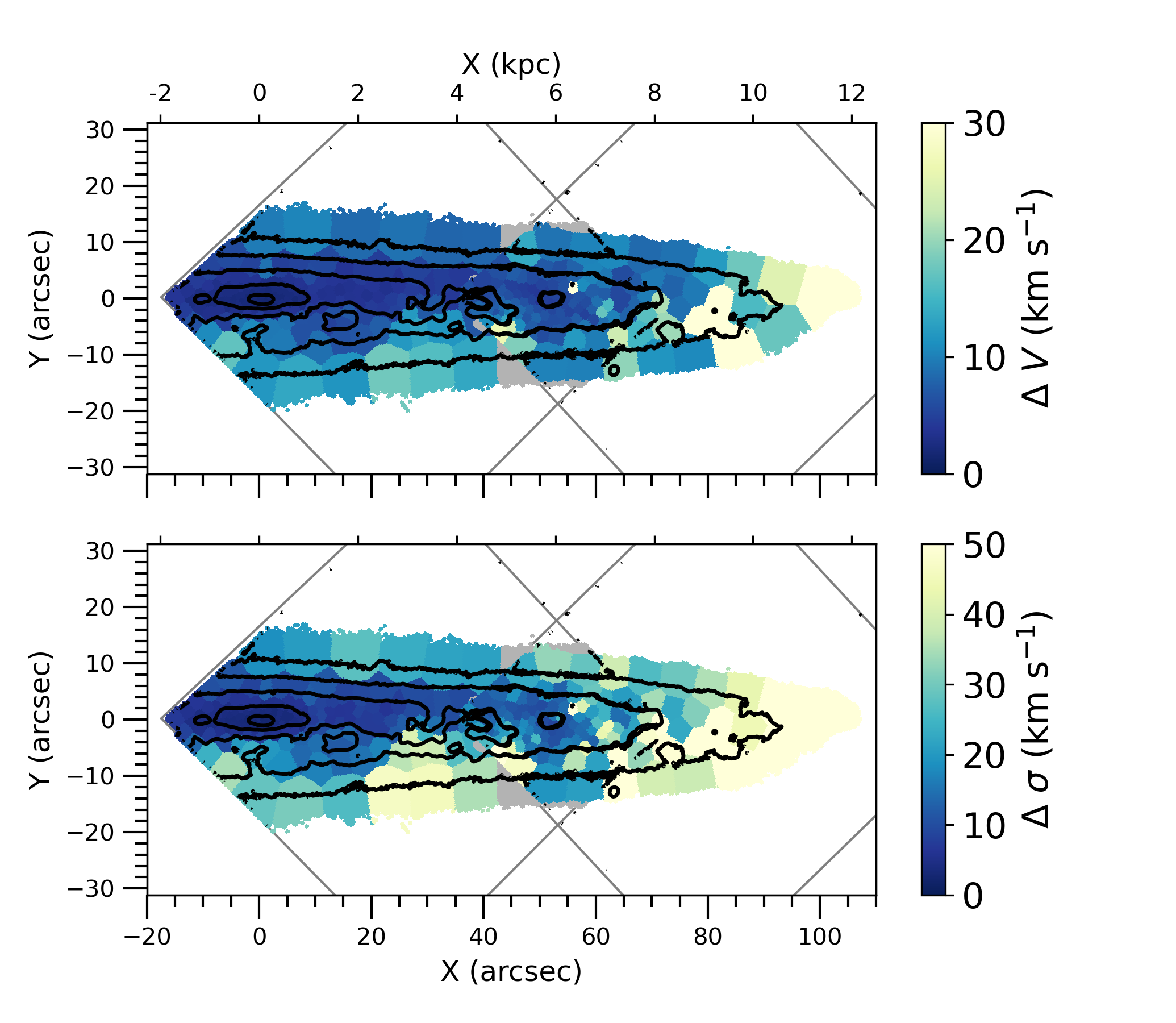}
	\end{minipage}
	\hfill{}
	\begin{minipage}[r]{\columnwidth}
		\centering
		\includegraphics[width=\textwidth]{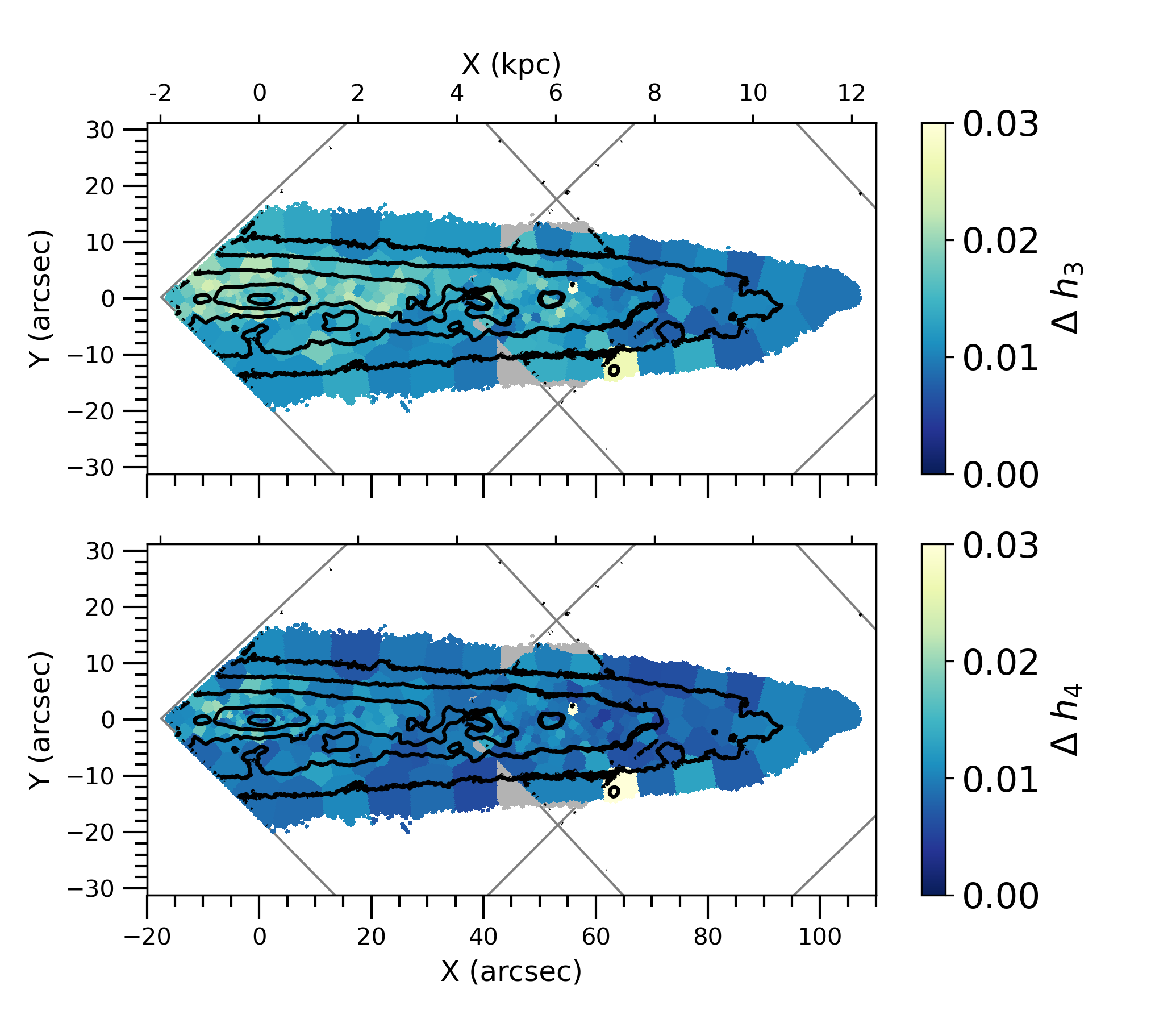}
	\end{minipage}
	\caption{Left: Uncertainty maps of the velocity and velocity dispersion obtained with Monte Carlo Simulations. Right: Uncertainty maps of $h_3$ and $h_4$ obtained with Monte Carlo Simulations. The approximate positions of the MUSE pointings are indicated by grey lines and the discarded bins are also displayed in grey, while the isophotes are displayed in black.
    }
	\label{fig:MC_kin}
\end{figure*}

The uncertainties of the stellar velocity are lowest in the central midplane and above ($\sim$~5~km~s$^{-1}$). In the Voronoi bins farthest away from the center, uncertainties reach $\sim$~30~km~s$^{-1}$.
The uncertainties of the stellar velocity dispersion show the same trend. The lowest uncertainties are found in the central region ($\sim$~10~km~s$^{-1}$) and the highest uncertainties in the outer parts ($\sim$~50~km~s$^{-1}$).
For $h_3$, we find uncertainties quite as high as the values for $h_3$ in the central midplane ($\sim$~0.02). Lower uncertainties ($\sim$~0.01) are found in the outer envelope, where we also see values of $h_3$ oscillating close to 0.
Also for $h_4$, uncertainties reach values of $\sim$~0.01, while the values of $h_4$ are mostly of the same order, supporting the penalization of higher moments at low signal-to-noise ratio by pPXF.

\begin{figure*}
	\begin{minipage}[l]{\columnwidth}
	    \includegraphics[width=\textwidth]{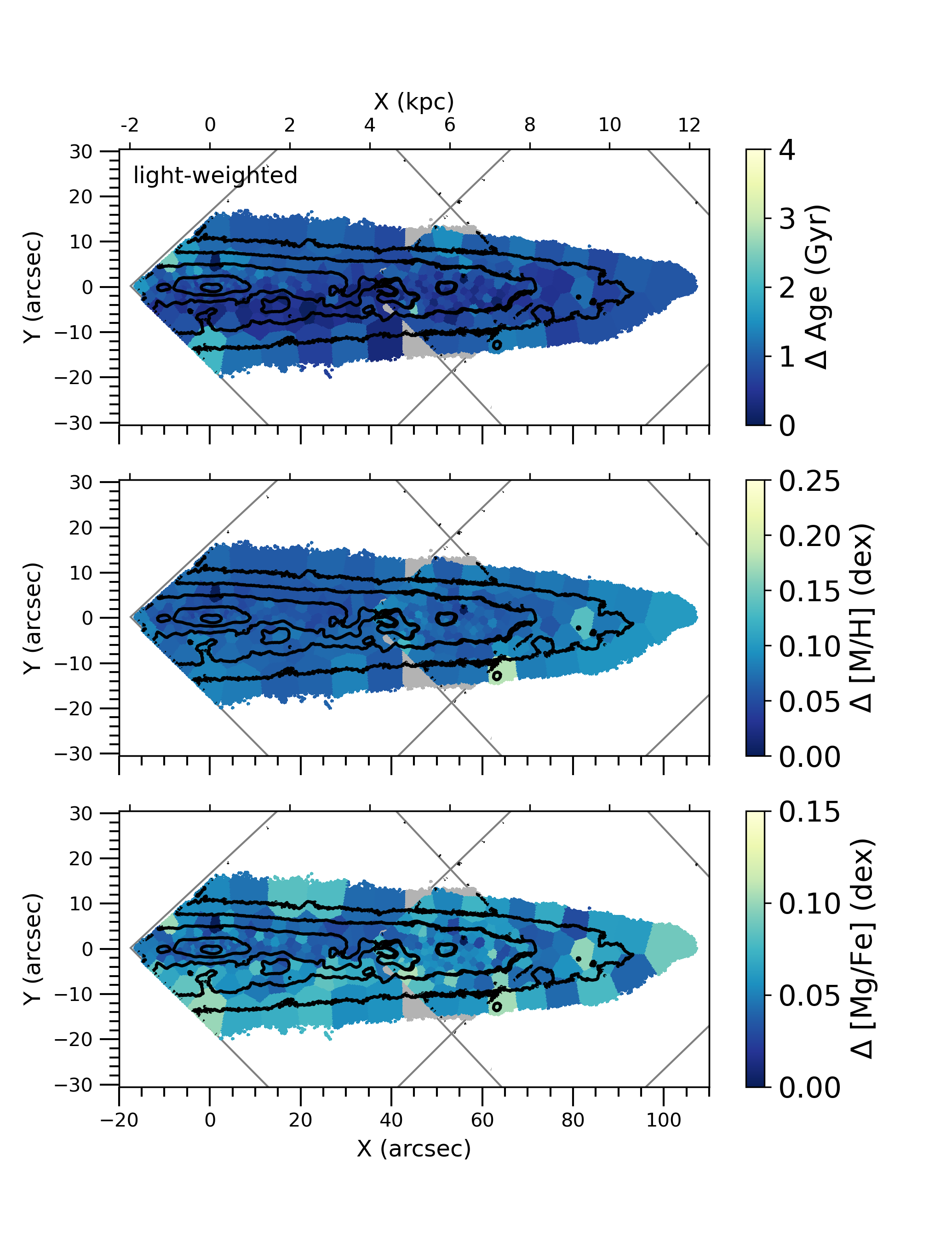}
	\end{minipage}
	\hfill{}
	\begin{minipage}[r]{\columnwidth}
		\centering
		\includegraphics[width=\textwidth]{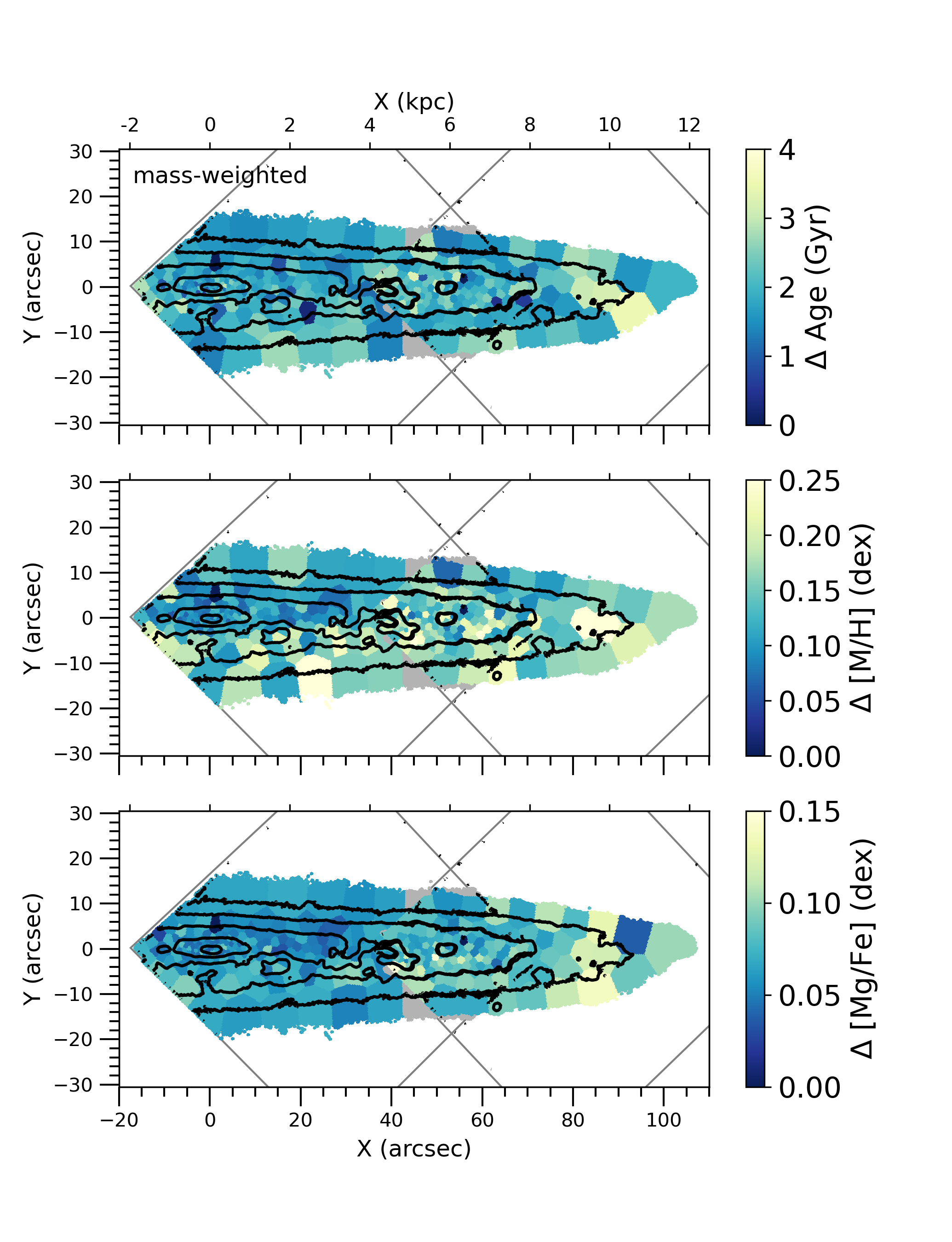}
	\end{minipage}
	\caption{Uncertainty maps of the light-weighted (left) and mass-weighted (right) stellar population parameters obtained with Monte Carlo Simulations. The approximate positions of the MUSE pointings are indicated by grey lines and the discarded bins are also displayed in grey, while the isophotes are displayed in black.
    }
	\label{fig:MC_pop}
\end{figure*}

From the uncertainty maps in Figure~\ref{fig:MC_pop}, we see for youngest ages lowest statistical uncertainties ($\sim$~0.5~Gyr in the midplane region of the light-weighted maps) and higher uncertainties for older ages on average ($\sim$~1~Gyr in the light-weighted maps and $\sim$~2.5~Gyr in the mass-weighted maps).
The uncertainties in metallicity [M/H] (Figure~\ref{fig:MC_pop}) are nearly constant ($\sim$~0.1~dex) over the whole galaxy half for the light-weighted results. For the mass-weighted results, uncertainties in metallicity [M/H] are lowest ($\sim$~0.1~dex) close to the central midplane and above and get larger ($\sim$~0.2~dex) towards the bottom of the galaxy and the outer parts.
For [Mg/Fe] (Figure~\ref{fig:MC_pop}), the uncertainties for the light-weighted results are lowest ($\sim$~0.05~dex) close to the central midplane and get slightly larger ($\sim$~0.1~dex) for the outer parts of the galaxy. For the mass-weighted results, uncertainties in [Mg/Fe] are lowest ($\sim$~0.05~dex) close to the central midplane, as well as above and below, and only get larger ($\sim$~0.1~dex) towards larger radii.

\begin{figure*}
	\begin{minipage}[l]{\columnwidth}
		\centering
		\includegraphics[width=\textwidth]{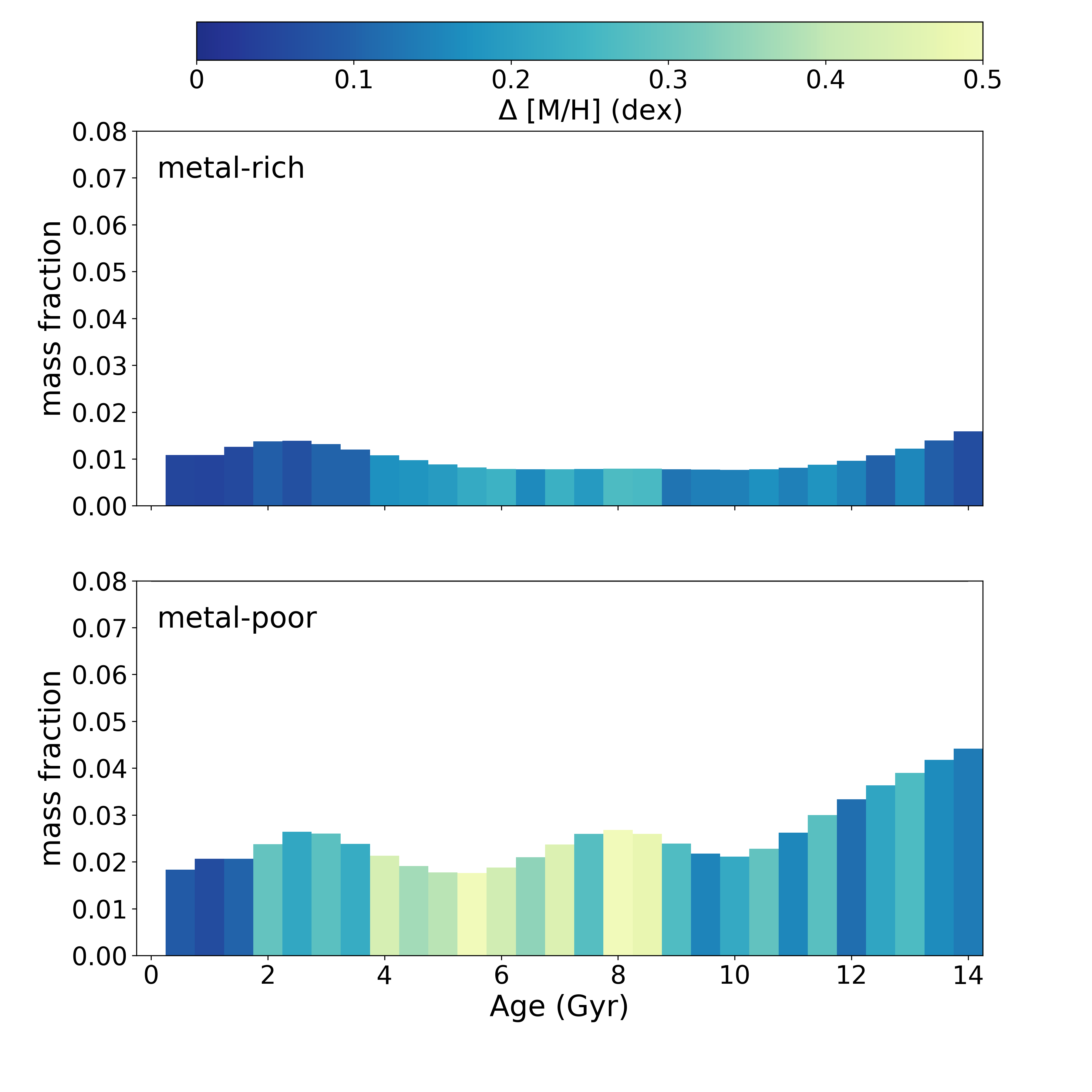}
	\end{minipage}
	\hfill{}
	\begin{minipage}[r]{\columnwidth}
		\centering
    	\includegraphics[width=\textwidth]{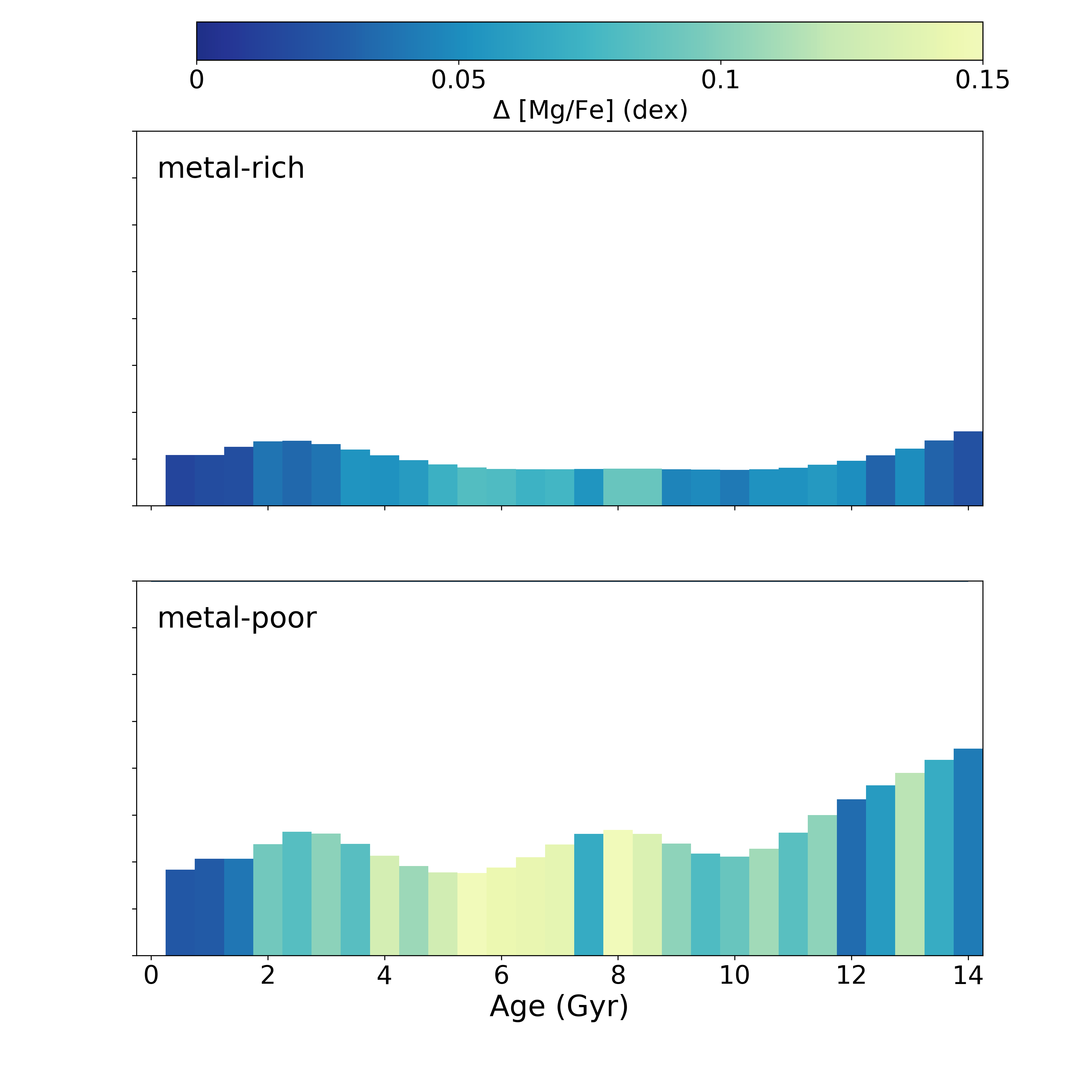}
	\end{minipage}
	\caption{Uncertainties of the star formation history obtained with Monte Carlo Simulations. Age bins are color-coded by the uncertainties of metallicity [M/H] (left), and by the uncertainties of [Mg/Fe] (right). The metal-rich component is always plotted at the top, whereas the metal-poor component is always plotted at the bottom.}
	\label{fig:MC_SFH}
\end{figure*}

\begin{figure}
    \centering
    \includegraphics[width=\columnwidth]{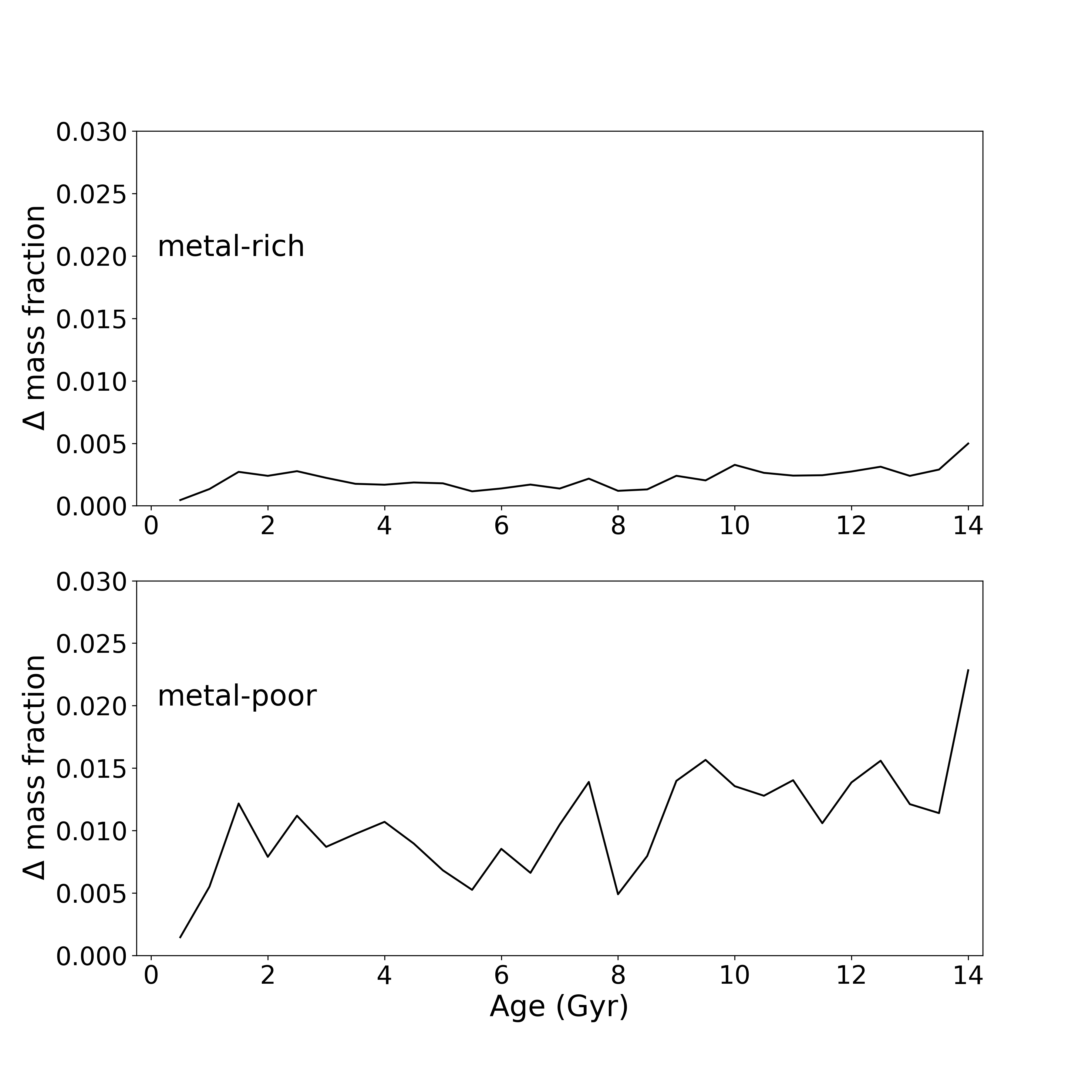}
    \caption{Uncertainties of the star formation history mass fractions obtained with Monte Carlo Simulations. The metal-rich component is plotted on the top, whereas the metal-poor component is plotted at the bottom.}
    \label{fig:MC_mf}
\end{figure}

Using the Monte Carlo simulations, performed in Section~\ref{subsec:uncertainties}, we calculated for each of the thousand realizations, the star formation history and computed the standard deviation from the mean metallicity [M/H] and [Mg/Fe] values as well as the mean mass fractions to obtain an estimate of the uncertainties. Since no regularization was used in the Monte Carlo simulations, the uncertainties do not show a smooth distribution. The uncertainties for metallicity [M/H], [Mg/Fe] and the mass fractions for the star formation history are shown in Figure~\ref{fig:MC_SFH} and \ref{fig:MC_mf}. 
Uncertainties in metallicity [M/H] and [Mg/Fe] (Figure~\ref{fig:MC_SFH}) as well as in mass-fractions (Figure~\ref{fig:MC_mf}) for the metal-rich component are in general lower than for the metal-poor component, reaching $\sim$~0.2~dex for [M/H], $\sim$~0.1~dex for [Mg/Fe] and $\sim$~0.003 for mass fraction. While the metal-poor component reaches higher uncertainties of $\sim$~0.5~dex for [M/H], $\sim$~0.15~dex for [Mg/Fe] and $\sim$~0.015 for mass fraction.
Moreover, we see lower uncertainties in metallicity [M/H] and [Mg/Fe] on average for the youngest and oldest ages in the star formation history, which tend to be more metal-rich and less enhanced in [Mg/Fe]. Those lower uncertainties reach $\sim$~0.1~dex ($\sim$~0.2~dex) for the metal-rich (metal-poor) component in [M/H] and $\sim$~0.05~dex (0.1~dex) for the metal-rich (metal-poor) component in [Mg/Fe].
The highest uncertainties in metallicity [M/H] and [Mg/Fe] are found for intermediate ages of 4 to 9~Gyr, where we find additionally lower metallicities and higher [Mg/Fe] abundances. Here, uncertainties reach $\sim$~0.2~dex (0.4~dex) for the metal-rich (metal-poor) component in [M/H] and $\sim$~0.1~dex (0.15~dex) for the metal-rich (metal-poor) component in [Mg/Fe].

\section{Systematic uncertainties}
\label{appendix:system_unc}

In our procedure, we only calculated statistical uncertainties from Monte-Carlo simulations (see e.g. Appendix~\ref{appendix:uncertainties}), which are much lower than the ones obtained in previous work using the same approach to analyze older galaxies  \citep{pinna_2019a,pinna_2019b,martig_2021}. 
Models of young stellar populations show larger differences, between spectra of different ages than older models for the same age difference. Thus for young galaxies adding some noise to the spectra as we do with Monte-Carlo simulations does not make pPXF change the choice of models. In young galaxies, systematic uncertainties may be dominating, but they are difficult to be quantified. We tested some of the potential sources of systematic uncertainties and we discuss them below.

To test the effect that our selection of specific MILES SSP models (Section~\ref{subsec:pop_models}) might have on the final results, we also fitted with pPXF all Voronoi binned spectra for stellar population parameters using the full range of ages in MILES SSP models. These include 9 values for metallicity between $-$1.26 to 0.4~dex, to still be in the safe range of metallicity for the SSP models, but now 53 values for age going from 0.03 to 14~Gyr with a resolution between 0.01 to 0.5~Gyr and the values for [Mg/Fe] of 0.0 and 0.4~dex. With this, the main features of the star formation history (higher metallicity at old stars and a drop in metallicity, higher metallicity and [Mg/Fe] at young ages, oscillations in [Mg/Fe] and the peaks in star formation around 3, 8, 14~Gyr) were still present. So the cuts we made in age did not affect our results significantly. 
It also needs to be considered, that all parameters like age, metallicity and [Mg/Fe]-abundance can be affected by the limited ranges of the SSP models. Other models might show different absolute values for the results, as discussed in several previous stellar population studies \citep[e.g.,][]{koleva_2008, baldwin_2017, ge_2018, ruiz-lara_2015}, while the vertical and horizontal gradients, as well as the spatial coherence, would be the same when using different models.

Secondly, we fitted with pPXF all Voronoi binned spectra for stellar population parameters using different values of regularization going from 1 to 100. With this, we found that the star formation history becomes smoother for higher values of regularization and the main features of the star formation history start to get lost using a regularization higher than 10. Choosing a higher regularization than 10 gives such a smoothed-out star formation history, that oscillations in metallicity and [Mg/Fe] as well as peaks in star formation are hardly visible. On the other hand, choosing a smaller value of regularization than 10 gives a noisier star formation history with larger jumps in metallicity and [Mg/Fe].
Our choice of regularization ensures that the systematics associated with it are minimized.

Another important point is the difficulty in retrieving the [$\alpha$/Fe] for young stars. 
For star-forming galaxies especially, the retrieval of $\alpha$-enhancement is more complicated, due to relevant absorption line features being significantly weaker in those galaxies consisting of young stellar populations, than for non-star-forming galaxies consisting of older stellar populations \citep[e.g.,][]{conroy_2014}.


\bsp	
\label{lastpage}
\end{document}